%% file: 00main.tex
\documentclass[acmsmall]{acmart}
\AtBeginDocument{%
  \providecommand\BibTeX{{%
    \normalfont B\kern-0.5em{\scshape i\kern-0.25em b}\kern-0.8em\TeX}}}

\setcopyright{acmcopyright}
\copyrightyear{0000}
\acmYear{2022}
\acmDOI{10.1145/3530814}

\acmJournal{CSUR}
\acmVolume{1}
\acmNumber{1}
\acmArticle{00}
\acmMonth{4}

\usepackage{subcaption}
\usepackage{booktabs}% http://ctan.org/pkg/booktabs
\usepackage{multirow}
\usepackage[flushleft]{threeparttable}
\usepackage{array}
\newcommand{\tabitem}{~~\llap{\textbullet}~~}
\usepackage{graphicx}
\usepackage{lscape}
\usepackage{enumitem}
\usepackage{graphicx}
\usepackage{textcomp} 
\usepackage{pifont}
\usepackage{adjustbox}
\usepackage{rotating}
\usepackage{makecell}
\usepackage{longtable}
\usepackage{supertabular}
\usepackage{verbdef}
\usepackage{xcolor}
\usepackage{colortbl}
\usepackage{ltablex,booktabs}
\definecolor{MyGreen}{RGB}{28,172,0}
\definecolor{MyRed}{RGB}{248,105,107}
\usepackage[toc,page]{appendix}
\usepackage{appendix}
\usepackage{xspace}
\begin{document}

\title{A Systematic Survey on Android API Usage for Data-Driven Analytics with Smartphones}

\author{Hansoo Lee}
\email{hansoolee@kaist.ac.kr}
\affiliation{%
  \institution{School of Computing, KAIST} 
  \streetaddress{291 Daehak-ro, Yuseong-gu}
  \city{Daejeon}
  \country{South Korea}
  \postcode{34141}
}
\author{Joonyoung Park}
\email{zelatore85@kaist.ac.kr}
\affiliation{%
  \institution{Graduate School of Knowledge Service Engineering, KAIST} 
  \streetaddress{291 Daehak-ro, Yuseong-gu}
  \city{Daejeon}
  \country{South Korea}
  \postcode{34141}
}
\author{Uichin Lee}
\authornote{The corresponding authors}
\email{uclee@kaist.ac.kr}
\affiliation{%
  \institution{School of Computing, KAIST} 
  \streetaddress{291 Daehak-ro, Yuseong-gu}
  \city{Daejeon}
  \country{South Korea}
  \postcode{34141}
}

\newcommand{\revision}[1]{{\color{black} #1}}
\renewcommand{\shortauthors}{H. Lee et al.}

\input{00abstract}



\maketitle
\input{01Introduction}
\input{02background_research}

\input{03methodology}
\input{04research_purpose_categorization}

\input{05categorization_of_android_mobile_usage_and_sensor_data}
\input{06results}
\input{07discussion}
\input{08conclusion}
\input{09appendix}

\begin{acks}
This research was supported by the Basic Science Research Program through the National Research Foundation (NRF) funded by the Korean government (MSIT) (2020R1A4A1018774, 2022R1A2C2011536).
\end{acks}

\bibliographystyle{ACM-Reference-Format}
\bibliography{bibliography.bib}


\end{document}

%% file: 00abstract.tex
\begin{abstract}
\revision{Recent industrial and academic research has focused on data-driven analytics with smartphones by collecting user interaction, context, and device systems data through Application Programming Interfaces (APIs) and sensors.} The Android OS provides various APIs to collect such mobile usage and sensor data for third-party developers. \revision{Usage Statistics API (US API) and Accessibility Service API (AS API) are representative Android APIs for collecting app usage data and are used for various research purposes as they can collect fine-grained interaction data (e.g., app usage history, user interaction type).} Furthermore, other sensor APIs help to collect a user's context and device state data, along with AS/US APIs. \revision{This review investigates mobile usage and sensor data-driven research using AS/US APIs, by categorizing the research purposes and the data types.} \revision{In this paper, the surveyed studies are classified as follows: five themes and 21 subthemes, and a four-layer hierarchical data classification structure.} This allows us to identify a data usage trend and derive insight into data collection according to research purposes. Several limitations and future research directions of mobile usage and sensor data-driven analytics research are discussed, including the impact of changes in the Android API versions on research, \revision{the privacy and data quality issues}, and the mitigation of reproducibility risks with standardized data typology.

\end{abstract}

%% file: 01Introduction.tex
\section{Introduction}

Currently, many people regard smartphones as an extension of their bodies, supporting everyday lives. According to recent statistics, 79\% of adults carry their mobile phones for 22 hours a day~\cite{stadd}, and users spend an average of 3 hours and 15 minutes a day~\cite{JoryMacKay}. Users interact with their smartphones an average of 2,617 times a day~\cite{dscout}. While holding or using a smartphone, fine-grained contextual information such as surrounding context, physical activity, and app usage patterns can be collected from built-in sensors and application programming interfaces (API). Thus, smartphones provide new opportunities to analyze a user's everyday life patterns. For example, mobile usage and sensor data serve as a valuable resource for the diagnosing and prognosis of mental and physical health~\cite{kourtis2019digital, liang2019survey}. 

\begin{figure*}
    \centering 
    \includegraphics[width=\textwidth]{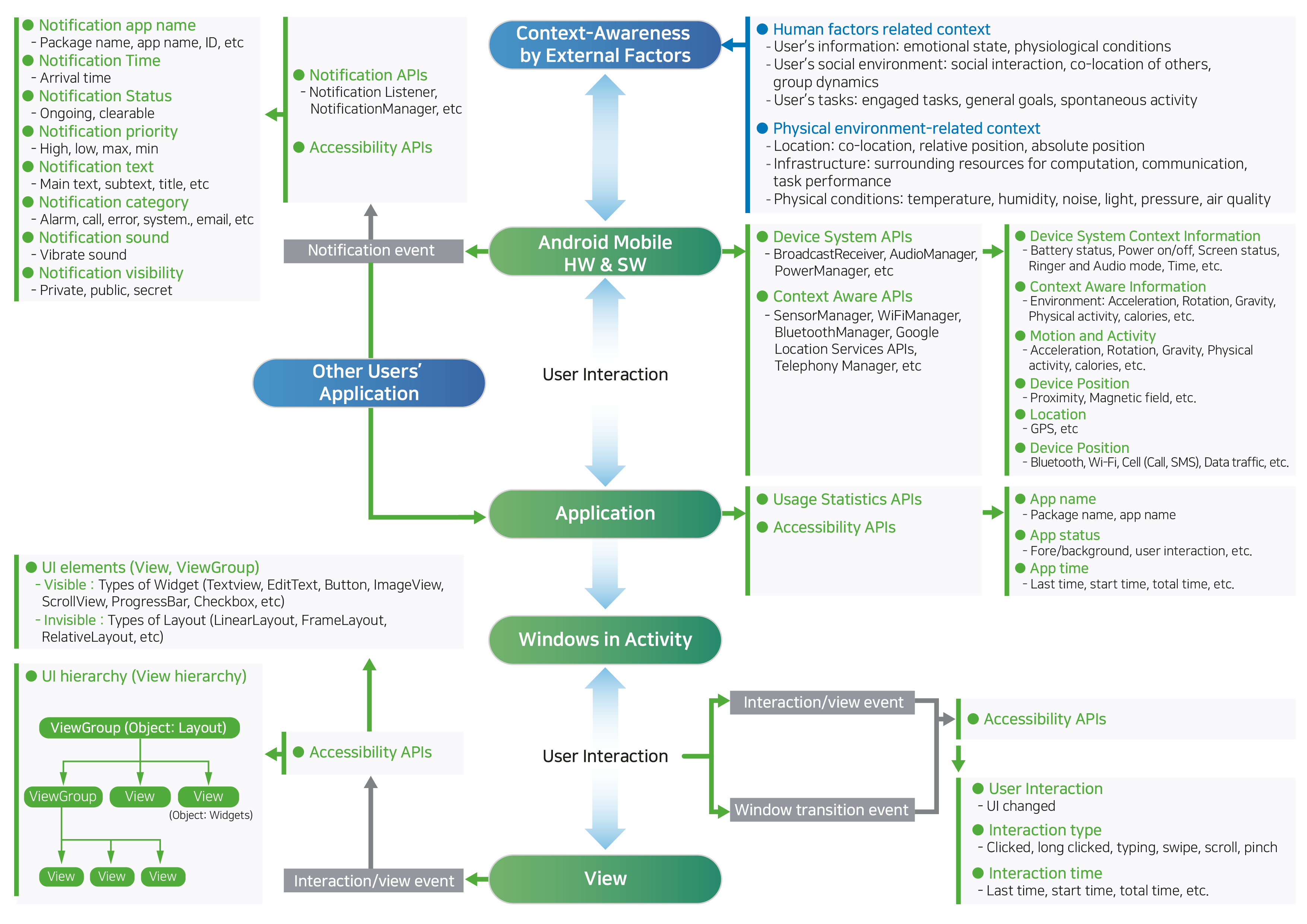} 
    \caption{Mobile usage and sensor data collection through Android APIs} 
    \label{fig:Interaction}
\end{figure*}

\revision{This review focuses on the Android mobile operating system (OS) which dominates 83.8\% of the global smartphone market~\cite{IDC:smartphone}.} As shown in Figure \ref{fig:Interaction}, Android built-in APIs can identify a user's life patterns based on mobility and activity data. Thus, we can collect interaction data (e.g., each app's touch event). Android OS allows developers to customize these built-in APIs to build applications for diverse passive data collection. For this reason, Android has been actively used in the research community. Researchers commonly use the Accessibility Service API (AS API) to collect user interaction and app usage pattern data among Android’s built-in APIs, also called Accessibility API (e.g., \emph{AccessibilityService} and \emph{AccessibilityEvent})~\cite{android.view.accessibility}. 
We can track app usage using various \emph{AccessibilityEvent} classes (e.g., transition type, view type, exploration type, and notification type)~\cite{AccessibilityService}. 

Another way to collect app usage pattern data is using the Android Usage Statistics APIs (US APIs)~\cite{Android.app.usage}. US API can be used to collect the app and device usage history and statistics. US API can access device usage history and statistics to obtain the currently running app information in third-party applications (Android API level 21 in 2014). Therefore, the US API has been mainly used to collect app usage pattern data in many studies (e.g.,~\cite{dutta2016introducing, singh2017usage, lee2018reducing, mehrotra2017understanding}). Beyond simple usage tracking, the AS API is still exclusively used to track the detailed user interface (UI) status information (e.g., UI changed status, interaction type, UI elements \& hierarchy, and notification) (e.g., \cite{chang2015investigating, kim2019understanding, lee2018click, yu2019understanding}).

Along with the AS/US APIs, other Android built-in sensors and APIs are also used to collect mobile usage and sensor data for the research. These data can be collected using Android built-in hardware (HW) sensors (e.g., accelerometer, gyro, GPS, Wi-Fi, Bluetooth, proximity, light, compass, pressure) and APIs (e.g., Sensor, Wi-Fi, Google Location Services APIs, Bluetooth APIs). In other words, we can use these Android built-in sensors and APIs to track a user’s physical activity, physiological state, and surrounding context information (e.g., location, ambient environment) for the research. Furthermore, Android system APIs (e.g., \emph{BroadcastReceiver}, \emph{BatteryManager}) can be used to collect device status information (e.g., battery information, screen status, device information setting, and ringer mode) (see Figure \ref{fig:Interaction}).

Leveraging data-driven analytics with smartphone data is widely performed in various research fields (e.g., sensors, human-computer interaction, ubiquitous computing, and mobile computing) as follows:
\begin{itemize}
\item Understanding a user's smartphone usage pattern~\cite{kim2019understanding, church2015understanding, lee2014hooked}
\item Identifying a user's notification checking factors, how the user is affected by the notification, and message management and scheduling~\cite{visuri2019understanding, pielot2014didn, chang2015investigating}.
\item Understanding a user's psychological \& health states and personal traits in digital phenotype and digital health research fields~\cite{bitsch2015psychologist, rooksby2019student, abdullah2014towards}.
\item Validating permission, improve authentication \& authorization and improve protection from malware in privacy \& security research fields~\cite{lau2014mimesis, fawaz2016privacy, ozcan2015babelcrypt}. 
\item Improving user interaction, accessibility, and optimization~\cite{chen2019messageontap, ryu2017siginterface, li2018appinite}.
\item Testing and programming support fields, such as graphical user interface (GUI) automated testing, crowdsourced system testing, programming support for developer, and app performance measurement~\cite{arruda2016capture, gu2019practical, muangsiri2017random}.
\item Development of frameworks and platforms to collect smartphone data~\cite{holzmann2017android, montague2015tinyblackbox}.
\end{itemize}

Table \ref{table:existingresearch1} lists the existing review studies investigating prior works that used various mobile usage and sensor data for specific research purposes. Rooksby et al.~\cite{rooksby2019student} reviewed existing studies that used smartphone sensor data to monitor student mental health. Kourtis et al.~\cite{kourtis2019digital} investigated the usage of sensor data by reviewing 62 studies that used mobile and wearable devices to determine digital biomarkers for Alzheimer's disease. Liang et al.~\cite{liang2019survey} reviewed 92 studies that used various types of data (e.g., online activities, blood pressure, ECG, EEG), including mobile and wearable device data, in a digital phenotyping study of mental health.

There are also review studies investigating research that used various mobile usage and sensor data under different experimental conditions or focused on specific mobile usage and sensor data as depicted in Table \ref{table:existingresearch1}. To gain a more generalizable understanding of mobile device use across different user populations, Church et al.~\cite{church2015understanding} analyzed the differences in data collected in 22 papers according to the number of different experimenters, experiment period, deployment, recruitment, incentives, and methods. Berkel et al.~\cite{van2017experience} investigated 110 studies using the experience sampling method (ESM) in experiment duration, types of triggers, response rate, compensation, personal device, and utilized data.

However, no prior studies performed a systematic survey of the existing research papers based on Android’s AS/US APIs and other built-in sensing APIs to the best of our knowledge. In addition, the existing review papers do not give a holistic view of mobile usage \& sensor data types for research purposes because they only investigated particular content (i.e., specific research purposes, data type, and experimental conditions). Furthermore, mobile usage and sensor data terminology are named differently, making it challenging to correctly identify mobile usage and sensor data items.
\input{tables/existingreviewstudies1}

This work first identifies the primary research purposes of those studies that use AS/US APIs. Then, this research reviewed the types of mobile usage and sensor data according to research purposes. Lastly, this research systematically reviewed the existing studies and categorized research purposes via a thematic analysis with affinity diagramming. In addition, this research categorized the collected data as follows: i.e., physical context sensing, system sensing, and interactive sensing data, by referring to API documents and prior studies.

The main contributions of this work can be summarized as follows: 
\begin{itemize}
\item This study provides an overview of the trends in existing research over the last ten years by categorizing \revision{the reviewed papers into several research themes through thematic analysis.} 
\item \revision{Next, this study reports on how to collect and use the data and summarizes the used data types according to surveyed studies and research purposes.}
\item Our findings help us to identify the relationships between research purposes and data types in mobile usage and sensor data-driven analytics research.
\revision{\item This study discusses challenges and insights (e.g., privacy and data quality issues, reproducibility issues of data term confusion, and research trends according to APIs release and update) that may arise when conducting various mobile usage and sensor data-driven analytics research.}
\end{itemize}

The rest of this article is organized as follows. Section \ref{sec:Background} provides background information on AS/US API, built-in sensors, and the other APIs. In Section \ref{sec:Methodology}, we present the overall process: finding, classifying, and selecting the literature covered for the systematic survey. In Section \ref{sec:research}, we classify the research purpose in surveyed studies. In Section \ref{sec:categorization}, we categorize the mobile usage and sensor data. In Section \ref{sec:datacategorization}, we present the results of data categorization. In Section \ref{sec:discussion}, we summarize the key findings from classification results, discuss the open issues (i.e., data term issues and practical issues in AS/US API use), privacy \& data quality issues, influence of API updates on research, list the limitations of this study and scope for future research. Finally, we conclude this article in Section \ref{sec:conclusion}.

%% file: tables/existingreviewstudies1.tex
% \usepackage{booktabs}
% \usepackage{graphicx}
\begin{table}[]
\centering
\caption{Papers including a summary of papers using mobile usage and sensor data}
\resizebox{\textwidth}{!}{%
\begin{tabular}{@{}lclll@{}}
\toprule
\multicolumn{1}{c}{\textbf{Ref.}} &
  \textbf{\begin{tabular}[c]{@{}c@{}}Number of Reviewed \\ Papers\end{tabular}} &
  \multicolumn{1}{c}{\textbf{Reviewed Items}} &
  \multicolumn{1}{c}{\textbf{Type of Investigated Data}} &
  \multicolumn{1}{c}{\textbf{Limitation}} \\ \midrule\midrule
Church et al. ~\cite{church2015understanding} &
  22 &
  \begin{tabular}[c]{@{}l@{}}Number of participants, experiment duration, \\ deployment, recruitment, incentives, method, \\ data collected, user input\end{tabular} &
  \begin{tabular}[c]{@{}l@{}}Time, communication, mobility, battery, \\ screen, notifications, multimedia, \\ applications, locations, touch\end{tabular} &
  \begin{tabular}[c]{@{}l@{}}For the 22 reviewed papers, only the experimental design method \\ (e.g., number of participants, period, recruitment, compensation) \\ was analyzed, and there is no classification of the research purpose, \\ so it is difficult to understand in detail what data was used for \\ what research purpose.\end{tabular} \\ \midrule
Rooksby et al. ~\cite{rooksby2019student} &
  19 &
  \begin{tabular}[c]{@{}l@{}}Utilized data, experiment duration, \\ number of participants\end{tabular} &
  \begin{tabular}[c]{@{}l@{}}Acceleration, activity, app usage, battery, charge, \\ browser history, Bluetooth, call logs, camera events, \\ creen, keyboard, UI, location, light sensor, microphone, \\ SMS/email\end{tabular} &
  \begin{tabular}[c]{@{}l@{}}In 19 studies, what kind of sensor data was used \\ was investigated, but only the research purpose \\ o/f a limited field (i.e., student mental health \\ monitoring in digital phenotyping) was investigated.\end{tabular} \\ \midrule
Kortis et al. ~\cite{kourtis2019digital} &
  62 &
  Utilized data, sense domain, metrics &
  \begin{tabular}[c]{@{}l@{}}Camera, microphone, acceleration/gyro, barometer, \\ touchscreen, geoposition, device use, ECG, PPG, \\ IR thermometer, ballistocardiography, galvanic skin response, \\ ambient light sensor, UV sensor, electromyogram (EMG)\end{tabular} &
  \begin{tabular}[c]{@{}l@{}}This research investigated 62 papers using mobile/wearable \\ sensor data only for specific research purposes \\ (i.e., Alzheimer's digital biomarker). \\ Furthermore, none of the 62 papers collected app usage data.\end{tabular} \\ \midrule
Berkel et al. ~\cite{van2017experience} &
  10 &
  \begin{tabular}[c]{@{}l@{}}Experiment duration, trigger, response rate, \\ compensate, personal device, utilized data\end{tabular} &
  \begin{tabular}[c]{@{}l@{}}ESM, location, phone calls, network related, bio-signal (s), \\ accelerometer, app usage, other (s)\end{tabular} &
  \begin{tabular}[c]{@{}l@{}}The 110 papers using a specific method (experience sampling \\ method on a mobile device) were reviewed only, \\ and the data used in each paper was briefly presented \\ without data categorization as the name of the device's sensor.\end{tabular} \\ \midrule
Liang et al. ~\cite{liang2019survey}&
  92 &
  \begin{tabular}[c]{@{}l@{}}\tabitem Data collection strategies\\ (data, strategy, pros, cons)\\ \tabitem List of affect lexicons \\ (category, lexicon, affect the type, vocabulary)\\ \tabitem Applications for mental health \\ (type, name, symptom, data)
  \\ \tabitem Affect recognition \\ (modality, category, work, algorithm)\\ \tabitem Behavior anomaly detection \\ (category, sensor, work, algorithm)\end{tabular} &
  \begin{tabular}[c]{@{}l@{}}\tabitem Physical data: facial expressions, behavioral data, \\ vocal data, context data (e.g., location, timestamp)\\ \tabitem Cyber data: online behaviors, social media data\\
  \tabitem Biological data: blood oxygen saturation, blood pressure, \\ heartbeat rate, respiration frequency, brain waves\\
  \tabitem Social data: call logs, text messages, GPS location, \\ hot spots, email, friendship list, online activities\\ (e.g., like, comment, sharing)\end{tabular} &
  \begin{tabular}[c]{@{}l@{}}Only 92 studies were reviewed for limited research purposes \\ (i.e., digital phenotyping of mental health), and the review \\ did not focus on papers using various mobile sensor and usage data.\end{tabular} \\ \bottomrule
\end{tabular}%
\label{table:existingresearch1}
}
\end{table}

%% file: 02background_research.tex
\section{Background}
\label{sec:Background}
This section describes Android built-in APIs used to collect mobile usage and sensor data described in Figure \ref{fig:Interaction}. Section \ref{sec:Android} explains the Android OS architecture and basic API framework (i.e., manager) for collecting mobile usage and sensor data. Section \ref{sec:Application} illustrates the APIs used to collect application interaction data such as user interaction and app usage patterns. Section \ref{sec:context} explains the APIs used to collect context-aware data (location, physical activity, ambient environment, network). Section \ref{sec:device} describes the APIs that can track device and system status (e.g., battery \& power status, screen on/off). Section \ref{sec:discussionAPIs} demonstrates why this study focuses on research using AS/US APIs among Android built-in APIs.

\subsection{Background on Android APIs to Collect Mobile Usage and Sensor Data}
\label{sec:Android}
This research describes the essential components and API framework to understand the Android APIs for mobile usage and sensor data collection. The Android Operating System (OS) consists of four layers which are \revision{the} application layer (system apps and third-party apps), Framework layer (API framework), Middleware layer (Native C/C++ library, Android runtime, Hardware Abstraction Layer), and kernel layer (Linux kernel) ~\cite{PlatformArchitecture, meng2018survey}. Third-party apps can access and utilize mobile usage and sensor data through the Android API framework's manager objects such as \emph{AccessibilityManager}, \emph{UsageStatsManager}, \emph{NotificationManager}, and \emph{SensorManager}, which are system-level services provided by Android. An instance of each object can be obtained through the \emph{getSystemService()} function~\cite{Context}, and specific data can be accessed via each instance. Table \ref{table:managers} shows constant parameter values in the process of obtaining an instance of each object and describes the Android API framework's manager objects that can be accessed through each instance.

Furthermore, there is an alternative way to collect mobile usage and sensor data besides accessing the manager in third-party apps. First, BroadcastReceiver can detect system status events and information (e.g., screen status, power status, battery status) occurring in Android OS. In addition, Bluetooth status information can be collected using only Bluetooth APIs, such as \emph{BluetoothAdapter} and \emph{BluetoothDevice}, without going through \emph{BluetoothManager}~\cite{Bluetoothoverview}. To collect location information, Google recommends that using the Google Location Services APIs is a more straightforward method for higher accuracy than the Location APIs that provide existing Android location-based and related services such as \emph{LocationManager} and \emph{LocationProvider}~\cite{android.location}. In some cases of UI event collection or third-party app development through AS API, \emph{AccessibilityService}, and \emph{AccessibilityEvent} are only used except for \emph{AccessabilityManager} (detailed explanation in Section \ref{sec:Application}). Through \emph{UsageStatsManager}, the researchers can collect usage history and statistics of apps and systems. Accordingly, \emph{UsageStatsManager} and \emph{AccessibilityService} are used more than other Managers and Services to identify smartphone usage patterns. First, the AS/US API can collect app usage and interaction, which are the core aspects of this review in Section \ref{sec:Application}. Second, context-awareness APIs (i.e., collecting ambient environments of the smartphone) and device \& system status APIs (i.e., collecting internal system and device status) that can be used in the study along with AS/US APIs are described in Sections \ref{sec:context} and Section \ref{sec:device}, respectively.
\input{tables/manager}

\subsection{Application Interaction Data Collection APIs}
\label{sec:Application}
As shown in Figure \ref{fig:Interaction}, we present an overview of the main functions, history, and collectible data of AS/US APIs. This overview aims to clarify the Android APIs used for app usage pattern data collection and derive the appropriate API terms for the literature search terms.

\subsubsection{Accessibility Service (AS) APIs}
Accessibility APIs are a set of Android APIs that are the basis for building applications to enhance user interfaces to support users with disabilities or users who are temporarily unable to fully interact with a device. Accessibility APIs include AccessibilityService and AccessibilityEvent. In this study, Accessibility APIs are abbreviated as ‘AS APIs.’ Android developer's documentation says that \emph{“the ability for you to build and deploy accessibility services was introduced with Android 1.6 (API Level 4) and received significant improvements with Android 4.0 (API Level 14)”}~\cite{Createyourownaccessibilityservice}. Further, \emph{“Accessibility services should only be used to assist users with disabilities (e.g., visually impaired people) in using Android devices and apps”} as per Android developer's documentation~\cite{AccessibilityService}. Besides, Google Android provides an application made with standard AS API (e.g., TalkBack), an assistive technology with functions such as screen touch, voice, Braille-based screen reader, speaker-based voice, and vibration~\cite{TalkBack}. For research purposes, AS API has been widely used to collect fine-grained data related to the user interaction data (e.g., types of interactions and app usage status), which provides useful information for behavior analysis.

The collection of fine-grained data (i.e., UI information via user interaction) is possible by grasping an \emph{AccessibilityEvent} delivered to the AS API through the \emph{onAccessibilityEvent()} callback method when the UI status is changed. The \emph{AccessibilityEvent} mainly appears when an event occurs in the UI due to touch manipulation, notification, or system change. First, to use the AS API, the service element should be included in the application element of the manifest file. The intent filter of the AS API must also be included in the service element. Next, the BIND\_ACCESSIBILITY\_SERVICE permission must be specified. The AS API can inform the system how and when to execute the AS API through a configuration variable set. For example, the developer can designate the \emph{AccessibilityEvent} type and package name to be processed by the AS API. Therefore, when UI events occur in \revision{the} Android system and app, the fine-grained data can be collected through AS API as follows: Types of touch and gesture interaction (e.g., click, long click, scroll, typing), time of interaction (e.g., interaction start/end time), view hierarchy construct, view elements, notification state change, and windows state change. As shown in Table A3 in Appendix C, the types of \emph{AccessibilityEvent} are categorized mainly into VIEW TYPES, TRANSITION TYPES, NOTIFICATION TYPES, EXPLORATION TYPES, and MISCELLANEOUS TYPES, for example, in  Android 11 (API level 30), and there are currently a total 47 types of \emph{AccessibilityEvent}~\cite{AccessibilityEvent}.

\subsubsection{Usage Statistics (US) APIs}
US API is provided mainly to obtain device and application usage history and statistics information~\cite{Android.app.usage}. The US API includes the representative APIs (e.g., \emph{UsageStatsManager}, \emph{UsageEvents}, \emph{StorageStatsManager}, \emph{ConfigurationStats}, and \emph{EventStats}) for accessing the app, as well as  device, network,  storage,  device configuration,  event type usage history and statistic~\cite{Android.app.usage}. US API has been widely used to analyze the application and device usage patterns along with AS API in many existing studies. Prior to Android 5.0 (API level 21), \emph{ActivityManager} was used to obtain information about a currently running foreground application. However, as the method of getting the currently running application information using \emph{ActivityManager}'s methods (e.g., \emph{getRunningTask()}, \emph{getRecentTasks()}, \emph{getRunningServices()}) have been deprecated since Android 5.0 (API level 21)~\cite{ActivityManager}. Instead of \emph{ActivityManger}, US API is used mainly to track device and application usage history and statistics information. Among the many classes in the US API above, we focus on the classes that can access app/device usage history and statistics such as \emph{UsageStatsManager}, \emph{UsageEvents}, \emph{UsageEvents.Event}\revision{~\cite{UsageStatsManager, UsageEvents, UsageEvents.Event}}. 

\emph{UsageStatsManager} can query application records. For example, by using the usage statistics constants (i.e., INTERVAL DAILY, INTERVAL WEEKLY, INTERVAL MONTHLY, and INTERVAL YEARLY), we can query the records to receive the classified results according to different periods (i.e., by the day, the week, the month, and the year)~\cite{UsageStatsManager}. Moreover, app usage time (e.g., last time used, app foreground/background used time), app usage status history (e.g., the status of foreground/background/user interaction), and package name can be retrieved using the \emph{UsageStats} object provided in the \emph{UsageEvents.Event} class through methods such as \emph{getPackageName()}, \emph{getLastTimeUsed()}, and \emph{getTotalTimeInForegroud()}. Furthermore, five types of bucket information (active, working set, frequent, rare, never) which define the usage status for each app according to how often the app was used recently can be obtained through \emph{UsageStatsManager} as of Android 10 (API level 29)~\cite{UsageStatsManager}. Various Android runtime information can be collected along with various application pattern information in US API. In addition, we can check the detailed information collected through \emph{UsageStatsManager}\revision{~\cite{UsageStatsManager} and \emph{UsageEvents.Event}~\cite{UsageEvents.Event}} in Table A4 in Appendix C.

\subsubsection{Notification API Framework}
Notification APIs refer to all APIs related to the collection of notification information that occurs by app usage or device status change in the smartphone. Representative Notification APIs include AS API, \emph{NotificationManager}, and \emph{NotificationListenerService}. \emph{AccessibilityEvent} in AS API has TYPE NOTIFICATION STATE CHANGED, which captures the notification information (e.g., time, app package name, displayed texts of the toasts such as small popup, and text information) notified through \emph{getEventType()}, \emph{getPackageName()}, \emph{getEventTime()}, \emph{getParcelableData()}, and \emph{getText()} methods. However, there is a limit to the information that can be accessed through AS API. \emph{NotificationListenerService} (launched Android API level 21) and \emph{NotificationManager} (launched Android API level 1) can collect the more specific notification information (e.g., notification app name, time, status, priority, text, category, sound, visibility). Currently, notification-related Android APIs such as \emph{NotificationListenerService} and \emph{NotificationManager} are widely used along with AS API to study notification usage behaviors in existing studies (e.g.,~\cite{kim2019understanding, mehrotra2017understanding, lee2019pass,  pradhan2017understanding, lee2018reducing, chang2017smartphone}).

\subsubsection{Call and Message APIs}
Call and Message API refers to APIs related to collecting app usage history/status of call and message (e.g., SMS, MMS), and sending/receiving notification information. Representative Call and Message APIs include AS/US API, \emph{TelephonyManager}, \emph{PhoneStateListener}, \emph{SmsManager}, and \emph{CallLog}. Existing studies have collected the call, short message service (SMS), multimedia message service (MMS) related app usage history/status, and sending/receiving notification information through AS/US API (e.g.,~\cite{pielot2014didn, anderson2019impact, chang2015investigating, lee2014hooked, welke2016differentiating, andone2016menthal, schweizer2014krakena, schweizer2014krakenb, dingler2017language}). However, AS/US API can access only the call/message-related app usage history and status information. Therefore, telephony APIs (e.g., \emph{TelephonyManager}, \emph{PhoneStateListener}, \emph{SmsManager}) or \emph{CallLog} can be alternatively used to access  fine-grained  information. \emph{TelephonyManager} can check the access and status of information about the device's telephony service (e.g., SMS, MMS, and call)~\cite{TelephonyManager}. In addition, we can use the methods of \emph{TelephonyManager} to check telephony services/status, access some types of subscriber information, and register listeners to receive phone status change notifications. Further, \emph{CallLog} can collect call-related data such as incoming/outcoming history information~\cite{Calllog}.

\subsection{Context-Awareness Android APIs} 
\label{sec:context}
Android APIs are mainly used for collecting context data as follows: motion, position, environmental context APIs, location context APIs, and network context APIs.

\subsubsection{Motion, Position, and Ambient Environment Context APIs} 
Representative Android APIs used to collect motion, position, and ambient environment context information include HW sensor-based sensor APIs such as \emph{SensorManager}, \emph{Sensor}, \emph{SensorEvent}, and \emph{SensorEventListener}~\cite{android.hardware}. HW sensors are specifically classified into motion detection sensors (e.g., accelerometers, magnetometers, gyroscope), physical position sensors (e.g., accelerometers, magnetometer, proximity), and ambient environment sensors (e.g., light, pressure, humidity, temperature)~\cite{SensorsOverview}. The context information based on mobile sensors and sensor APIs is collected and utilized in many existing studies (e.g.,~\cite{chang2017smartphone, pradhan2017understanding, park2017don, holzmann2017android, dingler2017language, ferreira2015aware}).

Besides, the \emph{Google Activity Recognition API} identifies the activity (i.e., walking, running, driving, standing still, cycling) is being performed by the user at each time with the sensors in the smartphone device through detecting a user's specific activity type constants (i.e., IN\_VEHICLE, ON\_BYCYLE, ON\_FOOT, RUNNING, STILL, TILTING, WALKING)\revision{~\cite{activityrecognition}}. Furthermore, the \emph{Google Activity Recognition Transition API} can detect a user's specific activity type constants (i.e., IN\_VEHICLE, ON\_BYCYLE, RUNNING, STILL, WALKING) to identify when a user starts or stops a specific activity~\cite{detectactivityandroid}. Existing studies used mobile sensors and Google Activity Recognition APIs to analyze the physical activity states (e.g.,~\cite{anderson2019impact, chang2015investigating, schweizer2014krakena, schweizer2014krakenb}).

\subsubsection{Location Context APIs}
Location APIs (e.g., \emph{LocationManager}, \emph{LocationProvider}) and \emph{Google Location Services API} are representative API frameworks mainly used to collect location context information in many existing studies (e.g., \cite{church2015understanding, kim2019understanding, welke2016differentiating}). In terms of efficiency and accuracy, \emph{Google Location Services APIs} are superior to the Location APIs, which can access the location information such as timestamp, latitude, longitude, altitude, accuracy, and speed of location. \cite{googlelocation}. If the \emph{Google Location Services API} is not available, the location information can be collected via the Location APIs (e.g., \emph{LocationProvider} and \emph{LocationManager}) in the traditional way as follows \cite{android.location}: (1) GPS location provider (GPS, AGPS), network location provider (GPS, Cell ID, Wi-Fi MAC ID), and the authority (android.permission.ACCESS\_FINE\_LOCATION or android.permission.ACCESS\_COARSE\_LOCATION) receive information from the network location provider (GPS, Cell ID, Wi-Fi MAC ID) and passive location provider (Cell ID, Wi-Fi MAC ID) are set in the manifest file~\cite{Locationmanager}. (2) After setting the authority, set the provider to be used by \emph{LocationManager}, and get updated location information from the GPS and network through the \emph{requestLocationUpdates()} method in the \emph{LocationManager}. It is also possible for a third-party app to collect the location provider and information (e.g., from the \emph{LocationManager} in real-time through \emph{LocationListener})~\cite{LocationListener}.

\subsubsection{Personal and Local Area Network Context APIs}
Smartphone network context information can be primarily divided into personal area network (PAN) and local area network (LAN) information.  Android's built-in Bluetooth sensor and APIs (including Bluetooth low energy; BLE) such as \emph{BluetoothManager}, \emph{BluetoothAdapter}, and \emph{BluetoothDevice} are mainly used to collect PAN usage data (e.g., device name, type, address, and connection status)~\cite{android.bluetooth}. 
Bluetooth Low Energy (BLE) scanning such as \emph{BluetoothLeScanner}~\cite{android.bluetooth.le} can be used to search for nearby BLE devices. Wireless LAN information can be collected using Android built-in network sensors (e.g., Wi-Fi) and network APIs such as \emph{WifiManager}  and \emph{ConnectivityManager}; e.g., Service Set Identifier (SSID), Basic Service Set Identifiers (BSSID), Received Signal Strength Indicator (RSSI), frequency, and the presence of connectable Wi-Fi networks can be detected in real-time (known as Wi-Fi scanning). \emph{ConnectivityManager} can be utilized to collect the network connection state information. In addition, \emph{NetworkStatsManager} and \emph{TrafficStats} can be used to obtain the network traffic usage statistics and history information (e.g., transmitted and received information of the network packets and bytes from all interfaces and mobile)~\cite{networktraffic, networkstatsmanager}.

\subsection{Device and System Status APIs}
\label{sec:device}
Android devise and system status information can be obtained by tracking events that occur according to real-time operation states (e.g., battery states, power on/off, network connection states) using the device and system status APIs. Currently, there are two main methods used to obtain device and system status information in Android OS. The first method is to register \emph{BroadcastReceiver} and selectively access device/system events using \emph{IntentFilter} to receive device/system status information and resources in Android~\cite{Broadcast}. When a specific device system status change occurs in the Android system, event information (e.g., ACTION\_AIRPLANE\_MODE\_CHANGED, ACTION\_BATTERY\_CHANGED/LOW/OKAY, ACTION\_BOOT\_COMPLETED) is delivered in the form of intent through the \emph{BroadcastReceiver}~\cite{Intent}. At this time, the third-party app can access the detailed device system information (e.g., battery statuses change, battery charge statuses such as low, high, system booting completion, bug reporting, phone connection, call button press, date change, system reboot, and device system status information). The second method is to access the manager (e.g., \emph{BatteryManager}, \emph{AlarmManager}, \emph{PowerManager}, \emph{AudioManager}) related to the device system by using the \emph{getSystemService()} method. For example, the \emph{BatteryManager} can broadcast all battery and charge details to a sticky Intent containing the state of charge without registering the \emph{BroadcastReceiver}~\cite{monitorthebattery}. Through \emph{BatteryManager}, third-party apps can access the device battery states information such as battery health (cold, dead, good, overheat, over-voltage), battery plugged (AC, USB, wireless), battery property (e.g., capacity, charge counter), and battery status (charging, discharging, full, not charging).

\subsection{Discussion of Android APIs}
\label{sec:discussionAPIs}

This section dealt with Android APIs for collecting mobile usage \revision{and sensor} data. \revision{To track various mobile usage and sensor data, the types of Android APIs were classified into three categories, and the data that each type of Android API can collect were investigated.} Among the Android APIs described above, AS/US APIs are the most representative APIs used in smartphone usage research targeting human behavior tracking by logging app usage, touch interactions, keystrokes, and notifications. Furthermore, most of the data collected by the context-aware Android APIs in Section {sec:context} and device \& status APIs in Section {sec:device} can be collected through AS/US APIs as well. Therefore, recent studies tended to collect app usage logs corresponding to user interaction, notification, and device \& system status, through AS/US APIs. Context-awareness APIs and device \& system status APIs were additionally used for further data collection, such as ambient environment (e.g., location, temperature, and humidity), physical activity state, network state (e.g., Bluetooths \& Wi-Fi status, data traffic), and detailed device \& system state changes (e.g., battery status, screen on/off, power on/off). For this reason, this research intends to focus on the studies that used AS/US API among various Android APIs.

%% file: tables/manager.tex
\begin{table}[]
\caption{Macro constants for service request according to Android Manager type ~\cite{Context}}
\resizebox{\textwidth}{!}{%
\begin{tabular}{@{}lll@{}}
\toprule
\multicolumn{1}{c}{\textbf{Constants: Requested Service}} & \multicolumn{1}{c}{\textbf{Managers: Returned Object}} & \multicolumn{1}{c}{\textbf{Description of Managers}} \\ \midrule\midrule
ACCESSIBILITY\_SERVICE                           & AccessibilityManager                         & Provides user feedback on UI events through registered event listeners. \\
USAGE\_STATS\_SERVICE                            & UsageStatsManager                            & Provides to the information of app and system usage history and statistics. \\
NOTIFICATION\_SERVICE                            & NotificationManager                          & Access to information for the notification events to users that happen in the system and app. \\
SENSOR\_SERVICE                                  & SensorManager                                & Access to the sensors related information in Android device. \\
TELEPHONY\_SERVICE                               & TelephonyManager                             & Access to information about telephone communication services. \\
LOCATION\_SERVICE                                & LocationManager                              & Access to information about location based services (LBS). \\
AUDIO\_SERVICE                                   & AudioManager                                 & Access to information about volume and ringer mode control.  \\       
BLUETOOTH\_SERVICE                               & BluetoothManager                             & Obtain the BluetoothAdapter resources and manage overall Bluetooth.   \\
WIFI\_SERVICE                                    & WIFIMananager                                & Manage the overall Wi-Fi network information.  \\ \bottomrule
\end{tabular}%
}
\label{table:managers}
\end{table}

%% file: 03methodology.tex
\section{Methodology}
\label{sec:Methodology}

\subsection{Literature Search}
This study created a keyword list and searched using those terms to find studies in which AS/US APIs were used on four scholarly literature search engines: Google Scholar\footnote{https://scholar.google.co.kr/}, Web of Science\footnote{http://www.webofknowledge.com/}, Scopus\footnote{https://www.scopus.com/}, ScienceDirect\footnote{https://www.sciencedirect.com/}, \revision{and ACM Digital Library.\footnote{https://dl.acm.org/}} Because AS/US APIs are used in various research fields, we searched for papers in the popular literature search engines~\cite{tober2011pubmed, gusenbauer2019google, martin2018google} as in prior review studies, in which papers from different fields can be found, instead of the web library databases of specific \revision{domains}. 

\begin{figure}[htp] 
    \centering 
    \includegraphics[width=0.7 \textwidth]{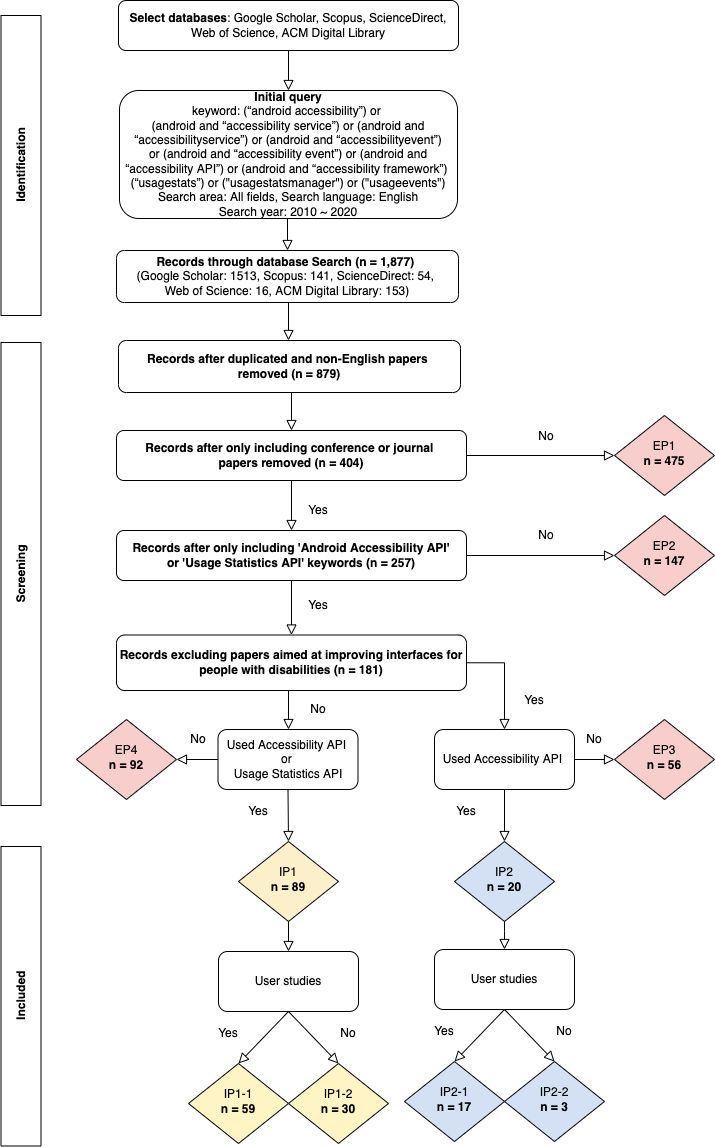} 
    \caption{Flow diagram for filtering and classification of target papers based on PRISMA} 
    \label{fig:Flowchart} 
\end{figure}

The keywords were divided to search papers into four lists: Android related, Accessibility APIs related, Usage Statistic API related, and combined keywords. This keyword list was composed of terms that are frequently used in \revision{the} existing literature. Table \ref{table:keywords} presents the list of the keywords used to search the papers. AS API became available in Android 1.6 (API level 4) in September 2009, and US API became available in the Android Framework from Android 5.0 (API level 21) in October 2014. Therefore, published papers were searched using AS/US APIs for the period from 2009 to 2020. The literature search keywords, period, language, and sites are summarized in Figure \ref{fig:Flowchart}.

\input{tables/keywords}

\subsection{Literature Filtering and Classification}
The studies were classified into two main categories: The studies were classified into two main categories: (1) studies on the application or framework development for collecting fine-grained datasets, and (2) studies on collecting fine-grained datasets using the AS/US API. As shown in Figure \ref{fig:Flowchart}, we used the customized Preferred Reporting Items for Systematic Reviews and Meta-Analyses (PRISMA) flow diagram~\cite{moher2009preferred} for our review to visualize the filtering and classification process for finding target papers. In addition, the classification results according to the four filtering criteria were divided into four categories of included papers (IP) to be reviewed and four categories of excluded papers (EP) to be excluded in our survey scope. 

Filtering criteria 1 studies corresponding to journals or proceedings, and filtering criteria 2 studies that mention the Android AS API or US API in the mobile environment. For filtering criteria 3 studies, we distinguished the existing studies using the AS to improve the interface accessibility of the users with disabilities (the original purpose of the Google Android policy) from papers that focused on AS API for other research purposes. Filtering criteria 4 studies used the AS or US API for generic research purposes beyond improving accessibility. According to four filtering criteria, only papers from academic journals or conferences written in English were included in the search results, and among them, papers using Android AS API or US API were selected.

The studies classified as IP1 and IP2 were considered to lie within the scope of the review. Through a review of these studies, we identified the research purpose of the studies, the types of AS/US APIs data used, and smartphone data collected via using other Android APIs. Table \ref{table:IP} shows the description and number of studies according to types of IP (IP1-1-AS, IP1-1-US, IP1-2-AS, IP1-2-US, IP2-1, IP2-2) which are the studies included in the scope of review. A total of 109 studies were investigated, as two studies~\cite{bonne2017exploring, ryu2017siginterface} were counted twice.

%% file: tables/keywords.tex
% \usepackage{booktabs}
% \usepackage{graphicx}
\begin{table}[]
\caption{List of Android, Accessibility APIs, Usage Statistics API, and combined keywords}

\resizebox{0.6\textwidth}{!}{%
\begin{tabular}{@{}ll@{}}
\toprule
\multicolumn{1}{c}{\textbf{List}} & \multicolumn{1}{c}{\textbf{Keywords}}                                                                                                                                                                                                                                                                                                   \\ \midrule\midrule
Android-related                   & “android"                                                                                                                                                                                                                                                                                                                               \\ \midrule
Accessibility APIs-related        & \begin{tabular}[c]{@{}l@{}}“accessibility"; “accessibility service"; “accessibilityservice"; \\ “accessibility API"; “accessibility framework"; \\ “accessibility event"; “accessibilityevent”\end{tabular}                                                                                                                                  \\ \midrule
Usage Statistic API-related      & “usagestats"; “usagestatsmanager"; “usageevent”                                                                                                                                                                                                                                                                                           \\ \midrule
Combined                          & \begin{tabular}[c]{@{}l@{}}“android" and "accessibility"; “android” and "accessibility service"; \\ “android” and "accessibilityservice"; “android” and “accessibilityevent"; \\ “android” and “accessibility event"; “android” and “accessibility API"; \\ “android” and “accessibility framework"; “android” and “usagestats”\end{tabular} \\ \bottomrule
\end{tabular}%
}
\label{table:keywords}
\end{table}

%% file: 04research_purpose_categorization.tex
\input{tables/IP}
\section{Research Purpose Categorization}
\label{sec:research}

We categorized the major research purposes of the included studies as IP1 (i.e., studies that used AS or US API \emph{other than} accessibility enhancement purposes for people with disabilities) in Section \ref{sec:categorizationIP1} and IP2 (i.e., studies that used AS or US API of accessibility enhancement purposes for people with disabilities) in Section \ref{sec:categorizationIP2} (See Table \ref{table:IP} for the detailed types of included paper). For categorization, we used affinity diagramming, a grouping technique that discovers meaningful rules among numerous data for classification according to the research purpose~\cite{holtzblatt2004rapid}. The process of categorizing the research objectives of several papers using the affinity diagramming technique is as follows. First, we extracted the words of related research purposes in the order of title, abstract, keyword, main text, and published site name of each paper. Among the words extracted from the reviewed papers, highly relevant words were grouped, and the words that were duplicated or unrelated to the purpose of the research were removed. Then, the research purposes representing the characteristics of the grouped words were found. The groups were then adjusted (e.g., groups of the same concept were merged, large groups were divided, small groups were merged with other groups) and named according to the research theme, and the themes were then divided into sub-themes. While reviewing the completed diagram from the beginning, we adjusted the themes and their labels.

\subsection{Categorization of IP1: Generic Research Work}
\label{sec:categorizationIP1}
We identify the research purposes of IP1 studies (a total of 89 papers for IP1), which were classified into five themes via affinity diagramming: (1) usage pattern (UP), (2) notification (NT), (3) enhancement of user interaction and experience (EUI/EUX), (4) privacy and security (PS), and (5) programming and testing support (PTS). Additionally, the five major themes were classified into 18 sub-themes, as shown in Table \ref{table:researchpurpose_IP1}.

\subsection{Categorization of IP2: Research Work for People with Disabilities}
\label{sec:categorizationIP2}
We identify and classify the research objectives of the IP2 studies that use the AS API for the original purpose (i.e., to improve the user interface to support the users with disabilities)~\cite{AccessibilityService}. IP2 can be included in the enhancement of UI and UX (EUI/UX) among the themes of the research purpose of surveyed studies corresponding to IP1. However, in the case of IP2, the sub-themes were classified differently than the IP1 sub-themes.
Accordingly, a total of 20 papers using the AS API for its original purpose have been primarily classified into three sub-themes.

\emph{\textbf{IFASD}} \emph{(Improvement of the Functionality of Accessibility Service for the users with Disabilities)}: 
IFASD studies were to improve the functions of the accessibility aspect of the existing AS API. They are classified further into two sub-research purposes. The first is to develop the accessibility service with improved functions in terms of accessibility enhancement by customized AS API~\cite{Createyourownaccessibilityservice}. The second is to develop improved functions to enhance the accessibility for users with disabilities by using the AS API and physical devices (e.g., braille, external sensor) together. In addition, the studies can be divided into three groups depending on the evaluation methods: (1) evaluating through user study by comparing and analyzing the existing AS API and the newly developed AS API in the paper, (2) evaluating only the developed AS API through the user study, and (3) not evaluating through the user study. Table \ref{table:disabled} summarizes studies in IFASD: what is proposed, what system or app is implemented compared to the basic Android AS API in terms of accessibility enhancement, and how usability tests were performed.

\emph{\textbf{DEASD}} \emph{(Development or Evaluation of an application using Accessibility Service for the users with Disabilities)}: 
DEASD studies were conducted for the purpose of developing an application program (e.g., calculator, configurable game, 3D printing) using the AS API rather than improving the limitations of the existing AS API. The research purpose of these papers is further divided into two subcategories. The first sub-theme is to enable TalkBack for verification of interactions with developed applications. In the second sub research purpose, AS API, such as TalkBack, is enabled and used not only for interaction purposes for the developed application programs but also for application development as its own AS API. Table \ref{table:disabled} summarizes studies in DEASD: what is the research purpose, what functions of the AS API were used, and how usability tests were performed about studies in DEASD.
\input{tables/researchpurpose_IP1}

\emph{\textbf{CFASD}} \emph{(Comparison of Functions of Accessibility Service for the users with Disabilities)}: In CFASD studies, AS API (e.g., TalkBack) is used to find the best interaction method for users with disabilities. The ultimate goal is to enhance smartphone use for users with disabilities. Most studies were conducted to solve the accessibility issues in smartphone utilization for visually impaired people. By contrast, Zhong et al.~\cite{zhong2015enhancing} and Rodriques et al.~\cite{rodrigues2015breaking} aimed to improve the accessibility for users with disabilities who have hand tremors or suffer from speech impairment. Table \ref{table:disabled} summarizes studies in CFASD: what is the research purpose, what functions of the AS API were used, and how usability tests were performed.

\revision{Further, according to the classification results of the research purposes in Sections \ref{sec:categorizationIP1} and \ref{sec:categorizationIP2}, we identified which papers belong to each research theme and sub-theme by IP type in Table \ref{table:researchpurpose}.} \revision{To summarize, this section classified research purposes with five themes and 21 sub-themes by the thematic analysis method. Through this classification, researchers can identify overall studies that belong to each research theme and sub-theme of existing research over the past 10 years.}
\input{tables/disabled}

\input{tables/researchpurpose}

%% file: tables/IP.tex
% \usepackage{booktabs}
% \usepackage{graphicx}
% \usepackage{booktabs}
% \usepackage{graphicx}
\begin{table}[]
\caption{Descriptions of meaning of IP}
\label{table:IP}
\centering
\begin{threeparttable}
\resizebox{\textwidth}{!}{%

\begin{tabular}{@{}llc@{}}

\toprule
\multicolumn{1}{c}{\textbf{Types of IP}} & \multicolumn{1}{c}{\textbf{Descriptions}}                                                                                                                                                                                  & \textbf{Number of Studies} \\ \midrule\midrule
IP1-1-AS                        & Research that perform user studies on AS API for improving the interface for the people with disability                                                                                                                                 & 43\tnote{*}                \\ \midrule
IP1-1-US                        & Research that perform user studies on US API                                                                                                                                                                                            & 17\tnote{*}                \\ \midrule
IP1-2-AS                        & \begin{tabular}[c]{@{}l@{}}Research that did not perform user studies to improve the interface \\ for the people with disability among the studies using AS API\end{tabular}                                      & 28\tnote{*}                \\ \midrule
IP1-2-US                        & Research that did not perform user study among studies using US API                                                                                                                                               & 3\tnote{*}                 \\ \midrule
IP2-1                           & \begin{tabular}[c]{@{}l@{}}Among the studies using the AS API, the studies where user studies were conducted \\ for the purpose of research to improve the interface for the people with disability\end{tabular}  & 17                \\ \midrule
IP2-2                           & \begin{tabular}[c]{@{}l@{}}Among the studies using the AS API, the studies where user study was not conducted \\ for the purpose of research to improve the interface for the people with disability\end{tabular} & 3                 \\ \midrule\midrule
\multicolumn{1}{c}{Overall}         &                                                                                                                                                                                                                   & 111-2\tnote{*} =109        \\ \bottomrule
\end{tabular}%
}
\begin{tablenotes}
\scriptsize
    \item[*]Bonné et al. [17] and Ryu et al. [131], which used both AS/US APIs, were included in IP1-1-AS/US and IP1-2-AS/US twice, respectively. \\ Therefore, the number of surveyed studies in this paper is 109, excluding the two out of a total of 111.
\end{tablenotes}
\end{threeparttable}
\end{table}

%% file: tables/researchpurpose_IP1.tex
%\label{table:researchpurpose_IP1}
%\caption{Research themes and sub-themes in IP1}
% \usepackage{booktabs}
% \usepackage{multirow}
% \usepackage{longtable}
% Note: It may be necessary to compile the document several times to get a multi-page table to line up properly
\begingroup % localize the following settings                                                       

\setlength\LTcapwidth{\textwidth} % default: 4in (rather less than \textwidth...)                                                                                                                           
\setlength\LTleft{0pt}            % default: \parindent                                                                                                                                                     
\setlength\LTright{0pt}           % default: \fill                                              
\scriptsize
\setlength{\tabcolsep}{5.0pt}
%\caption{Research themes and sub-themes in IP1}
\begin{tabularx}{\textwidth}{lll}
\caption{Research theme and sub-theme in IP1}\\
\toprule
\multicolumn{1}{c}{Theme} &
  \multicolumn{1}{c}{Sub-Theme} &
  \multicolumn{1}{c}{Description} \\ \hline\hline
\endfirsthead
\multicolumn{3}{c}%
{{\bfseries Table \thetable\ continued from previous page}} \\
\hline
\multicolumn{1}{c}{Theme} &
  \multicolumn{1}{c}{Sub Theme} &
  \multicolumn{1}{c}{Descrition} \\ \hline
\endhead
\multirow{3}{*}{\raisebox{-2.9em}{Usage Pattern (UP)}} &
  Generic Usage Pattern Analysis (GUPA) &
  \begin{tabular}[c]{@{}l@{}}Research on the analysis of generic mobile usage\\ patterns based on mobile data-driven analytics \\ according to specific conditions (e.g., society, culture, \\ education, health status, age group) of users\end{tabular} \\ \cline{2-3} 
 &
  \begin{tabular}[c]{@{}l@{}}Machine Learning-based Model \\ and System (MLMS)\end{tabular} &
  \begin{tabular}[c]{@{}l@{}}Research on machine learning models and \\ system development to understand \\ user characteristics based on the mobile usage pattern\end{tabular} \\ \cline{2-3} 
 &
  Framework and System Development (FSD) &
  \begin{tabular}[c]{@{}l@{}}Research on developing the framework and\\ the system using AS/US APIs, and other Android APIs \\ for enabling fine-grained mobile data collection\end{tabular} \\ \hline
\multirow{3}{*}{\raisebox{-3.4em}{Notification (NT)}} &
  \begin{tabular}[c]{@{}l@{}}Identifying Factors responding \\ to Notifications (IFN)\end{tabular} &
  \begin{tabular}[c]{@{}l@{}}Research to understand or predict how users react to and \\ interact with which notifications in what context\end{tabular} \\ \cline{2-3} 
 &
  \begin{tabular}[c]{@{}l@{}}Notification Timing and \\ Notification Management (NT/NM)\end{tabular} &
  \begin{tabular}[c]{@{}l@{}}Screening research to reduce interruption by notification, \\ just in time notification scheduling research, and research\\ on notification management according to the importance\end{tabular} \\ \cline{2-3} 
 &
  How Notification Affects a User (HNAU) &
  \begin{tabular}[c]{@{}l@{}}Research to identify how notifications affect users \\ (e.g., attentiveness, emotion, action, responsiveness, \\ interruptible, learning)\end{tabular} \\ \hline
\multirow{3}{*}{\raisebox{-3.0em}{\begin{tabular}[c]{@{}l@{}}Enhancement of \\ UI and UX (EUI/UX)\end{tabular}}} &
  Computational Enhancement (CE) &
  \begin{tabular}[c]{@{}l@{}}Research for user interaction and experience improvement \\ (e.g., upgrading system performance, lip recognition) \\ through computational enhancement (e.g., automation, \\ recognition, offloading, microservice)\end{tabular} \\ \cline{2-3} 
 &
  User Interface Enhancement (UIE) &
  \begin{tabular}[c]{@{}l@{}}Research for improving UI (e.g., personalized \\ conversation interface, conversation-aware interface, \\ verification by password-less access, multi-modal interface)\end{tabular} \\ \cline{2-3} 
 &
  User Interface Profiling (UIP) &
  \begin{tabular}[c]{@{}l@{}}Research on UI profiling such as complexity measurement of \\ the UI or power management of CPU/GPU\end{tabular} \\ \hline
\multirow{4}{*}{\raisebox{-3.6em}{\begin{tabular}[c]{@{}l@{}}Privacy and \\ Security (PS)\end{tabular}}} &
  Authentication System/Scheme (ASS) &
  \begin{tabular}[c]{@{}l@{}}Research on the suggestion of systems/schemes for \\ user authentication in the Android mobile environment\end{tabular} \\ \cline{2-3} 
 &
  Access and Permission Control (APC) &
  \begin{tabular}[c]{@{}l@{}}Research to identify user intention and context for user \\ access and permission control or to suggest new methods\end{tabular} \\ \cline{2-3} 
 &
  Privacy-Preserving System (PPS) &
  \begin{tabular}[c]{@{}l@{}}Research to propose a system for preserving \\ user privacy in Android mobile environment\end{tabular} \\ \cline{2-3} 
 &
  \begin{tabular}[c]{@{}l@{}}Monitoring and Detection for \\ Privacy and Security (MDPS)\end{tabular} &
  \begin{tabular}[c]{@{}l@{}}Research to detect and monitor spying behavior and \\ hidden attacks in terms of personal privacy and \\ security protection\end{tabular} \\ \hline
\multirow{5}{*}{\raisebox{-3.8em}{\begin{tabular}[c]{@{}l@{}}Programming and \\ Testing Support (PTS)\end{tabular}}} &
  GUI Automated Testing (GAT) &
  \begin{tabular}[c]{@{}l@{}}Research to propose an automated model \\ and system for GUI testing of Android applications\end{tabular} \\ \cline{2-3} 
 &
  Crowdsourced Testing (CT) &
  Research to test Android applications through crowdsourcing \\ \cline{2-3} 
 &
  Automated Programming Support (APS) &
  \begin{tabular}[c]{@{}l@{}}Research on supporting the automation of \\ programming tasks in system development\end{tabular} \\ \cline{2-3} 
 &
  \begin{tabular}[c]{@{}l@{}}Performance Measurement \\ and Management (PMM)\end{tabular} &
  \begin{tabular}[c]{@{}l@{}}Research to measure or manage the \\ performance of Android devices and systems\end{tabular} \\ \cline{2-3} 
 &
  \begin{tabular}[c]{@{}l@{}}Test Recorder and Script \\ Generator Technique (TRSGT)\end{tabular} &
  \begin{tabular}[c]{@{}l@{}}Research on technology to automatically generate test scripts \\ by converting and encoding test actions into test \\ script format via user interactions for automated testing\end{tabular} \\ \hline

\label{table:researchpurpose_IP1}
\end{tabularx}

\endgroup

%% file: tables/disabled.tex
% \usepackage{booktabs}
% \usepackage{multirow}
% \usepackage{graphicx}

\begin{table}[]
\caption{Analysis of research in IP2: (1) what is proposed, (2) Accessibility Service function used or improved, (3) usability test method according to the research purpose  (IFASD, DEASD, and CFASD)}
\resizebox{\textwidth}{!}{%

\begin{tabular}{@{}clllccl@{}}
\toprule
                        & \multicolumn{1}{c}{}                       & \multicolumn{1}{c}{}                                                                                                                                                                                                                                      & \multicolumn{1}{c}{}                 
& \multicolumn{3}{c}{\textbf{Methods of Usability Tests}}                                                                                                                                                                                                                                                                                                                                                         \\ \cmidrule(l){5-7} 
\multirow{-2}{*}{\textbf{Theme}} & \multicolumn{1}{c}{\multirow{-2}{*}{\textbf{Ref.}}} & \multicolumn{1}{c}{\multirow{-2}{*}{\textbf{What is Proposed}}}                                                                                                                                                                                                    & \multicolumn{1}{c}{\multirow{-2}{*}{\textbf{\begin{tabular}[c]{@{}c@{}}Implemented Functions or Utilized \\ Functions in Accessibility Service\end{tabular}}}}                                                      & \textbf{\begin{tabular}[c]{@{}c@{}}Number of \\ Participants\end{tabular}} & \textbf{\begin{tabular}[c]{@{}c@{}}Disabilities of\\ Participants\end{tabular}}                     & \multicolumn{1}{c}{\textbf{Evaluation Metrics}}                                                                                                                                                                                                \\ \midrule\midrule
                        & Chinchole et al. \cite{chinchole2017artificial}                    & \begin{tabular}[c]{@{}l@{}}A low-cost auxiliary development device \\ with economical efficiency, high efficiency, mobility, \\ and ease of use via AI and detection sensors\end{tabular}                                       & \begin{tabular}[c]{@{}l@{}}AI-based image recognition, collision \\ detection, and obstacle detection \\ functions to navigate the environment\end{tabular}                                                & Uncounted                                                         & Non-disabled                                                                             & \multicolumn{1}{c}{-}                                                                                                                                                                                                                \\ \cmidrule(l){2-7} 
                        & Oliveira et al. \cite{de2018application}                     & \begin{tabular}[c]{@{}l@{}}An improvement of voice command to improve \\ the interaction for visually impaired people\end{tabular}                                                         & \begin{tabular}[c]{@{}l@{}}Voice commands improvement and \\used Talkback to evaluation \end{tabular}                                                                                                                                                                                  & 6                                                                 & Visually impaired                                                                          & \begin{tabular}[c]{@{}l@{}}System Usability Scale ~\cite{brooke1996sus}, setup time duration,\\qualitative questions autonomy, satisfaction,\\daily use, fatigue, and feeling of being lost\end{tabular}\\ \cmidrule(l){2-7} 
                        & Rodriques et al. ~\cite{rodrigues2015breaking}                   & \begin{tabular}[c]{@{}l@{}}Macros to enhance time spent \\ searching and inaccessible content with visually impaired\end{tabular}                                   & \begin{tabular}[c]{@{}l@{}}Manual macro creation, automatic macro \\ generation, and macro production to perform \\ multiple steps of interaction with a single tab\end{tabular}                           & 1                                                                 & \begin{tabular}[c]{@{}c@{}}Visually impaired, \\speech disorder, \\motor impaired\end{tabular} & Interviewed user experience                                                                                                                                                                                                          \\ \cmidrule(l){2-7} 
                        & Rodriques et al. ~\cite{rodrigues2018aidme}                    & A tutorial guide for the Accessibility Service                                                                                                                                                                                      & \begin{tabular}[c]{@{}l@{}}Interactive non-visual tutorials that \\ support audio recording and launch tutorials\end{tabular}                                                                              & 11                                                                & Visually impaired                                                                          & Success rate of tasks                                                                                                                                                                                                                \\ \cmidrule(l){2-7} 
                        \multirow{-1}{*}{IFASD} & Trindade et al. ~\cite{trindade2018hybrid}                     & \begin{tabular}[c]{@{}l@{}}A Hybrid-Brailler that integrate physical \\ and touch screen devices to provide fast, accurate, \\ and flexible nonvisual input for visually impaired users\end{tabular}                                           & \begin{tabular}[c]{@{}l@{}}Input solution with combining physical \\ Braille keyboard and touch screen devices\end{tabular}                                                                                & 11                                                                & Visually impaired                                                                          & \begin{tabular}[c]{@{}l@{}}Words per min, typing error rate, editing ratio, \\ task completion time\end{tabular}                                                                                                                     \\ \cmidrule(l){2-7} 
                        & Zhang et al. ~\cite{zhang2017interaction}                        & \begin{tabular}[c]{@{}l@{}}Interaction proxies to runtime repair \\ and accessibility improvement of app.\end{tabular}                                                                                                                       & Interaction proxies                                                                                                                                                                                        & 14                                                                & \begin{tabular}[c]{@{}c@{}}Low vision, \\ visually impaired\end{tabular}                   & \begin{tabular}[c]{@{}l@{}}Accessibility barriers, user experience, usefulness \\ and potential of proposed method based on interview\end{tabular}                                                                                   \\ \cmidrule(l){2-7} 
                        & Zhang et al. ~\cite{zhang2018robust}                        & \begin{tabular}[c]{@{}l@{}}Robust annotations for app interface element \\ to solve the access problem of \\ the users with disabilities in Accessibility Service \end{tabular}     & \begin{tabular}[c]{@{}l@{}}Capture tool, template screen tool, \\ runtime library, alternative collection \\ and annotation tools to get properties\\ of each element from Accessibility Node\end{tabular} & 5, 1                                                              & \begin{tabular}[c]{@{}c@{}}Developer, \\ visually impaired\end{tabular}                    & \begin{tabular}[c]{@{}l@{}}- Screen equivalence heuristics: \\ the error rate of misjudgment of two screens\\ - Runtime repair based interview\end{tabular}                                                                          \\ \cmidrule(l){2-7} 
                        & Zhong et al. ~\cite{zhong2014justspeak}                      & \begin{tabular}[c]{@{}l@{}}A voice control to use with other Android apps \\ and talkback to support the system quickly \end{tabular}                                                               & \begin{tabular}[c]{@{}l@{}}Hands-free voice control \\ based on Accessibility Service\end{tabular}                                                                                                         & -                                                                 &                                                                                            & \multicolumn{1}{c}{-}                                                                                                                                                                                                                \\ \cmidrule(l){2-7} 
 & Zhong et al. ~\cite{zhong2015enhancing}                       & \begin{tabular}[c]{@{}l@{}}The system-wide assistance service \\ that can solve the accessibility problem of \\ people with hand tremors\end{tabular}                                                & \begin{tabular}[c]{@{}l@{}}Touch area enhancement in \\ Accessibility Service\end{tabular}                                                                                                                  & 8                                                                 & Motor-impaired                                                                             & Acquisition time via types of touch and trial interfaces                                                                                                                                                                             \\ \midrule
                        & Alnfiai et al. ~\cite{alnfiai2017brailletap}                      & The BrailleTap calculator using tap gestures                                                                                                                                                                                                     & Talkback's screen reader                                                                                                                                                                                   & 2                                                                 & Visually impaired                                                                          & Average time spent for calculation the equation                                                                                                                                                                          \\ \cmidrule(l){2-7} 
                        & Chen et al. ~\cite{chen2018construction}                         & \begin{tabular}[c]{@{}l@{}}The voice assistant for fall detection, safety care, \\ phone's accessibility, daily broadcasting information, \\ view description via AI and mobile computing\end{tabular} & Used the Accessibility Service                                                                                                                                                                             & \multicolumn{1}{l}{}                                              & \begin{tabular}[c]{@{}c@{}}The volunteers \\ with "wearing a \\ blinder"\end{tabular}      & \begin{tabular}[c]{@{}l@{}}"Computing time of object detection in \\ different internet environment"\end{tabular}                                                                                                                    \\ \cmidrule(l){2-7} 
                        & Correa et al. ~\cite{correa2018development}                      & A configurable game for education                                                                                                                                                                                                              & \begin{tabular}[c]{@{}l@{}}Different types of touch input\\ (Tap, Double tap, swipe, tow finger swipe)\end{tabular}                                                                                        & 6                                                                 & Visually impaired                                                                          & \begin{tabular}[c]{@{}l@{}}Efficiency and satisfaction\end{tabular}                                                                           \\ \cmidrule(l){2-7} 
                        & Samonte et al. ~\cite{samonte2019braille3d}                      & \begin{tabular}[c]{@{}l@{}}A learning app incorporating \\ haptic and a voice feedback for braille recognition \\ and 3D printing for visually impaired students\end{tabular}                              & \begin{tabular}[c]{@{}l@{}}Talkback's screen reader and\\ user feedback (haptic, vibration)\end{tabular}                                                                                                   & Uncounted                                                         & Visually impaired                                                                          & Success rate of tasks execution                                                                                                                                                                                                      \\ \cmidrule(l){2-7} 
                      \multirow{-3}{*}{DEASD}  & Ghidini et al. ~\cite{ghidini2016developing}                      & \begin{tabular}[c]{@{}l@{}}A electronic calendar by voice, keyboard, \\ and touch command interaction, and identify \\ what interaction is best for visually impaired\end{tabular}                                                 & Talkback's screen reader                                                                                                                                                                                   & 4                                                                 & Visually impaired                                                                          & \begin{tabular}[c]{@{}l@{}}Interviewed user experience when participant perform \\ the proposed tasks\end{tabular}                                                                                                                   \\ \cmidrule(l){2-7} 
                        & Hann et al. ~\cite{hann2013best}                         & \begin{tabular}[c]{@{}l@{}}A system of digital accessible information \\ for print disabled people \end{tabular}                                                                               &  Used and Customized Talkback                                                                                                                                                                                                          & -                                                                 & \multicolumn{1}{l}{}                                                                       & \multicolumn{1}{c}{-}                                                                                                                                                                                                                \\ \cmidrule(l){2-7} 
                        & Ganz et al. ~\cite{ganz2014percept}                         & \begin{tabular}[c]{@{}l@{}}A low cost and a friendly indoor \\ navigation system for visually impaired\end{tabular}                                                                                                                          & \begin{tabular}[c]{@{}l@{}}Used Accessibility Service's functions:\\ "explore by touch" and "hovering"\end{tabular}                                                                                        & Uncounted                                                         & Not visually impaired
                                                                             & \begin{tabular}[c]{@{}l@{}}Correctness of the navigation instructions and \\ the user interface by performing certain functions\end{tabular}                                                                                         \\ \cmidrule(l){2-7} 
 & Pareddy et al. ~\cite{pareddy2019x}                      & \begin{tabular}[c]{@{}l@{}}A system that captures and embeds the \\ meaning of basic contents as images for visually impaired.\end{tabular}                                                                                       & \begin{tabular}[c]{@{}l@{}}Used Accessibility Service \\ to collect UI hierarchy\end{tabular}                                                                                                              & 5                                                                 & Visually impaired                                                                          & \begin{tabular}[c]{@{}l@{}}Survey of X-Ray's learnability, comfort, usefulness, \\ perceived speed, perceived accuracy, \\ satisfaction with Likert scale of 1 to 7\end{tabular}                                                     \\ \midrule
                        & Dobosz et al. ~\cite{dobosz2016control}                       & \begin{tabular}[c]{@{}l@{}}Finding the best method among existing user interface \\ approaches for playing mobile games\end{tabular}                                                                                                                         & Talkback's screen reader                                                                                                                                                                                   & 11                                                                & Visually impaired                                                                          & \begin{tabular}[c]{@{}l@{}}Interface reaction time, commands correctness, \\ screen size influence when using mobile game\end{tabular}                                                                                               \\ \cmidrule(l){2-7} 
                      \multirow{-1}{*}{CFASD}  & Rodriques et al. ~\cite{rodrigues2015getting}                   & \begin{tabular}[c]{@{}l@{}}A method to investigate smartphone adoption \\ and usage by beginner users with visual impairments\end{tabular}                                                                                                     & \begin{tabular}[c]{@{}l@{}}Used Talkback to understand device usage \\ and interaction observation\end{tabular}                                                                                            & 5                                                                 & Visually impaired                                                                          & \begin{tabular}[c]{@{}l@{}}Interviewed the main concerns, observations, challenges, \\ barriers, and experiences about smartphones, \\ device usage, and interaction observation\end{tabular}                    \\ \cmidrule(l){2-7} 
 & Alshyban et al. ~\cite{alshayban2020accessibility}                     & \begin{tabular}[c]{@{}l@{}}Tools to evaluate Android accessibility's \\ capabilities in terms of apps, developers, and users\end{tabular}                                                                                                    & \begin{tabular}[c]{@{}l@{}}Used the Accessibility Service to monitor the \\ AccessibilityEvent when each app is crawled\end{tabular}                                                                       & 66                                                                & Practitioners                                                                              & \begin{tabular}[c]{@{}l@{}}Conduct survey of current practices, challenges, \\ accessibility tools, and guidelines related to accessibility\end{tabular}                                                                             \\ \bottomrule
\end{tabular}%
\label{table:disabled}
}
\end{table}

%% file: tables/researchpurpose.tex
% \usepackage{booktabs}
% \usepackage{multirow}
% \usepackage{graphicx}
\begin{table}[]
\caption{Results of research purpose categorization via IP types}
% \usepackage{booktabs}
% \usepackage{multirow}
% \usepackage{graphicx}
\resizebox{\textwidth}{!}{%
\begin{tabular}{@{}llllllll@{}}
\toprule
\multicolumn{1}{c}{\textbf{Theme}}               
& \multicolumn{1}{c}{\textbf{Sub Theme}}                                                    
& \multicolumn{1}{c}{\textbf{Ref: IP1-1-AS}}                                                                                                                               
& \multicolumn{1}{c}{\textbf{Ref: IP1-1-US}}                            
& \multicolumn{1}{c}{\textbf{Ref: IP1-2-AS}}        
& \multicolumn{1}{c}{\textbf{Ref: IP1-2-US}} 
& \multicolumn{1}{c}{\textbf{Ref: IP2-1}} 
& \multicolumn{1}{c}{\textbf{Ref: IP2-2}} 
\\ \midrule\midrule
\multirow[b]{3}{*}{\raisebox{0.6em}{UP}}  

& GUPA                            
& ~\cite{blanke2014mining, church2015understanding, kim2019understanding, lee2014hooked, ferreira2014contextual}                                  
& ~\cite{yuan2019much, qin2018deciphering, qin2019association, radesky2020young, singh2017usage, welke2016differentiating} 
&                                                                         &
&
&\\ \cmidrule(l){2-8} 
& \begin{tabular}[c]{@{}l@{}}MLMS\end{tabular}    
& ~\cite{lee2018click}                                              
& ~\cite{khan2020personal, dutta2016introducing, yan2019bridging}    
& ~\cite{yu2019understanding}                                         
&
&
&\\ \cmidrule(l){2-8} 
& \begin{tabular}[c]{@{}l@{}}FSD\end{tabular}               
& ~\cite{andone2016menthal, zheng2017cleaning}                      
&                                 
& ~\cite{bitsch2015psychologist, holzmann2017android, montague2015tinyblackbox, schweizer2014krakena, schweizer2014krakenb, ferreira2015aware}  
& 
&
&\\ \midrule
\multirow[b]{3}{*}{\raisebox{0.6em}{NT}}   
& \begin{tabular}[c]{@{}l@{}}IFN\end{tabular}            & ~\cite{pielot2014didn, visuri2019understanding, chang2015investigating} 
& ~\cite{mehrotra2017understanding}                                       &                                  
&
&
&\\ \cmidrule(l){2-8} 
& \begin{tabular}[c]{@{}l@{}}NT/NM\end{tabular}    
& ~\cite{lee2018exploring, okoshi2015attelia, okoshi2016towards, park2017don, pielot2017beyond, pradhan2017understanding}                                               
& ~\cite{lee2018reducing, lee2019pass}                                                                       &                                                                                                            &
&
&\\ \cmidrule(l){2-8} 
& HNAU                                                         
& ~\cite{anderson2019impact, chang2017smartphone, dingler2017language, komuro2017relationship, lee2019does, pielot2014situ} 
&                                                                                                            &                                                                                                            &                                                  
&&\\ \midrule
\multirow[b]{6}{*}{\raisebox{1.6em}{EUI/UX}}       
& CE                                                                
& ~\cite{sun2018lip}                                                                                        
&                                                                                                       
& ~\cite{elazhary2018w, tarakji2018voice, wang2018client}                                                   
&                                                  
&&
\\ \cmidrule(l){2-8} 
& UIE                                                               
& ~\cite{chen2019messageontap, li2018appinite, zhang2017interaction}                
&  
& ~\cite{rauen2018empowering, ryu2017siginterface}                        
& ~\cite{ryu2017siginterface}    &&                         
\\ \cmidrule(l){2-8} 
& UIP                                                                 
& ~\cite{lin2019user, riegler2018measuring}                        
&                                        
& ~\cite{riegler2015ui}                       
&&& 
\\ \cmidrule(l){2-8} 
& \begin{tabular}[c]{@{}l@{}}IFASD \end{tabular}                                    
&                       
&                                        
&                       
& 
&~\cite{de2018application, rodrigues2015breaking, rodrigues2018aidme, trindade2018hybrid, zhang2017interaction, zhang2018robust, zhong2015enhancing}
& ~\cite{chinchole2017artificial, zhong2014justspeak}
\\ \cmidrule(l){2-8} 
& \begin{tabular}[c]{@{}l@{}} DEASD \end{tabular}                                                        
&                        
&                                        
&                      
&
& ~\cite{alnfiai2017brailletap, chen2018construction, correa2018development, samonte2019braille3d, ghidini2016developing, ganz2014percept, pareddy2019x}
& ~\cite{hann2013best}
\\ \cmidrule(l){2-8} 
& \begin{tabular}[c]{@{}l@{}} CFASD \end{tabular}                                                   
&                       
&                                        
&                       
&
& ~\cite{dobosz2016control, rodrigues2015getting, alshayban2020accessibility}
&
\\ \midrule

\multirow[b]{4}{*}{\raisebox{1em}{PS}}            
& ASS                                                              
& ~\cite{aras2019multilock, kraus2017use, canfora2016silent}                                                 & ~\cite{zhu2019riskcog, zhu2020hybrid, torres2019behavioral}  
&                                                                                                            &                                                  
&&
\\ \cmidrule(l){2-8} 
& APC                                                             
& ~\cite{bonne2017exploring, rahman2018iac}                                  
& ~\cite{bonne2017exploring}                                         
&                        
&                                                  
&&\\ \cmidrule(l){2-8} 
& PPS                                                                
& ~\cite{fawaz2016privacy, lau2014mimesis, ozcan2015babelcrypt}          
& ~\cite{andriotis2020allow}
& ~\cite{fernandes2016appstract}         
&                                                  
&&\\ \cmidrule(l){2-8} 
& \begin{tabular}[c]{@{}l@{}}MDPS\end{tabular} & ~\cite{naseri2019accessileaks}                
&                                            
& ~\cite{rastogi2016these, shao2018understanding, li2018chatting, vishwamitra2017mcdefender, leguesse2020reducing}                         
& ~\cite{wang2019dcdroid, wang2020identifying} &&\\ \midrule
\multirow[b]{5}{*}{\raisebox{1,2em}{\begin{tabular}[c]{@{}l@{}}PTS\end{tabular}}} & GAT                                                                     
& ~\cite{arruda2016capture}                                                                   
&                                                                                                            & ~\cite{muangsiri2017random, san2016gui, gu2019practical}                                 
&                                                  \\ \cmidrule(l){2-8} 
& CT                                                                    
&                                                                         &                                                                                     
& ~\cite{lian2018cat, wu2017appcheck}         
&                     &&                             \\ \cmidrule(l){2-8} 
& APS                                                            
& ~\cite{li2017programming, li2020interactive}
&                                                 
& ~\cite{li2018kite, li2017sugilite}             
&                                                  \\ \cmidrule(l){2-8} 
& \begin{tabular}[c]{@{}l@{}}PMM\end{tabular}                    & ~\cite{bissig2018towards}                                       
& ~\cite{xiang2020dynamical}                          
&                                                                                                           &                         &&                         \\ \cmidrule(l){2-8} 
& \begin{tabular}[c]{@{}l@{}}TRSGT\end{tabular}            & ~\cite{fazzini2017barista}                             
&                                           & ~\cite{liu2017mechanism, negara2019practical}                 &                                    &&
\\ \bottomrule
\end{tabular}%
}
\label{table:researchpurpose}
\end{table}

%% file: 05categorization_of_android_mobile_usage_and_sensor_data.tex
\section{Categorization of Android Mobile Usage and Sensor Data}
\label{sec:categorization}
We investigate the mobile usage and sensor data that Android APIs can collect. Moreover, we present the type of mobile usage and sensor data used for each of the research purposes of the surveyed papers classified in Section \ref{sec:categorizationIP1}. 

\subsection{Excluded Studies in Data Categorization}
The data in IP1 studies (i.e., \revision{used AS or US API \emph{other than} accessibility enhancement purposes}) is divided into three cases as follows: (1) mobile usage and sensor data collected through user studies, (2) mobile usage and sensor data collected through other methods such as developer test, crawling, crowdsourcing other than user study, (3) types of mobile usage and sensor data that can be collected in a framework developed for data collection purposes. We included all studies in the data categorization, except for one study~\cite{san2016gui}, for which it was difficult to determine the scope of the data used. Next, among the surveyed studies corresponding to IP2 (i.e., \revision{used AS or US API for accessibility enhancement purposes}), only eight surveyed studies~\cite{alshayban2020accessibility, chen2018construction, rodrigues2015getting, rodrigues2015breaking, rodrigues2018aidme, pareddy2019x, zhang2018robust, zhong2015enhancing} that performed user studies and collected data are included in the data categorization. Therefore, 96 studies out of the total number of surveyed studies (n = 109) are included in data categorization.
\begin{figure*}
    \centering 
    \includegraphics[width=0.9\textwidth]{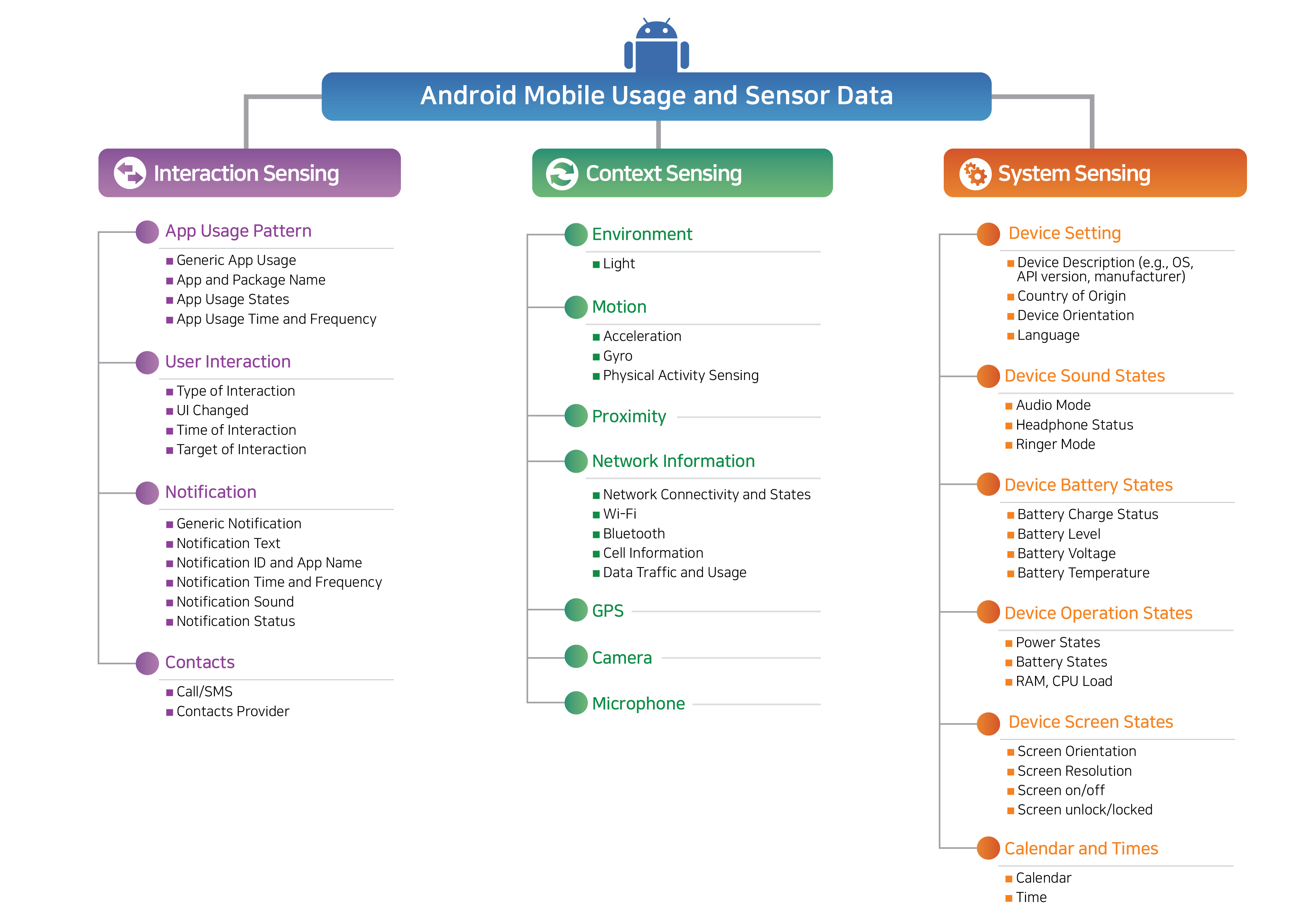} 
    \caption{Android mobile usage and sensor data categories} 
    \label{fig:datacategorization}
\end{figure*}

\subsection{Method of Data Item Categorization}
As shown in Figure \ref{fig:datacategorization}, we create a hierarchical structure of the mobile usage and sensor data item terms using data categorization to identify and classify data collected from the third-party Android apps developed through Android APIs (e.g., AS/US APIs) in the surveyed studies. The hierarchical structure of the mobile usage and sensor data terms are created in four phases: (1) find and extract the data term list from papers, (2) create a prototype of data categorization, (3) rearrange the data categorization, and (4) reflect on issues in the categories. 

\input{tables/existingdatacategory}

\subsubsection{Find \& Extract Data Term} \label{sec:find} 
A data term list of mobile usage and sensor data that can be collected through the Android framework/application developed in the prior studies is obtained. Data terms described in studies were individual words or words included in categories and sentences. Examples of each description type for the screen on/off data are as follows:
\begin{itemize}
\item Word form: “Screen (on, off)”~\cite{anderson2019impact}, “Screen on/off”~\cite{chang2015investigating}, “Screen on/off event”~\cite{church2015understanding}, “Screen state”~\cite{komuro2017relationship}
\item Sentence form: “Time when the phone’s screen was turned on or off”~\cite{pielot2014didn}, “Determine whether the user is currently using the phone”~\cite{park2017don}
\end{itemize}

\subsubsection{Devising Data Typology} We create a typology of data by conducting data term integration and making the hierarchical structure for the prototype of data categorization. The data term lists are grouped based on similarities. Unified data terms are defined as the representative data term for each grouped data by referring to the data term of the existing research (Table \ref{table:existingdatacategorization}) and Android developer documentation \cite{AndroidDeveloperdocumentation}. In addition, a hierarchical prototype of data categorization with five layers was created by referring to the data categorization of the existing research (Table \ref{table:existingdatacategorization}) and the Android developer documentation \cite{AndroidDeveloperdocumentation}.

\subsubsection{Revising Data Typology} \label{sec:rearranging} We revised the hierarchical structure when carefully reviewing the overall data categorization. The steps are as follows: (1) The data are grouped into categories or the existing categories are divided into multiple categories, reflecting on issues such as difficulties in understanding the collected mobile usage and sensor data term and in showing all data types on the chart because of too many data types)); (2) the categories of data terms not described in the papers are excluded; (3) if the papers state that data are collected or can be collected, but no categories of data in the prototype of data categorization are available, a new category is created and added in the prototype of data categorization.

Among the surveyed studies, there was a paper that stated "sensors" without specifying what sensor data is for the data type collected. We cover this issue in more detail by classifying the surveyed studies based on the degree of easy understanding of the data topology in Section \ref{sec:description}.

Figure \ref{fig:datacategorization} shows the data categorization-based four-layer hierarchical structure results according to the unification of the mobile usage and sensor data terms described in the studies. The meaning of the second layer is as follows: 
\begin{itemize}
\item \emph{\textbf{Interaction sensing}}: Mobile usage and sensor data that users generate by interacting with or initiating an app or system service (excluding system-specific data)
\item \emph{\textbf{Context sensing}}: Mobile usage and sensor data to understand the \revision{system and user context}
\item \emph{\textbf{System sensing}}: Mobile usage and sensor data on the main state of the system (e.g., resources, screen status, device settings)
\end{itemize}

\revision{To summarize, this section presented the standardized terms and classification with a four-layer hierarchical structure, which helps researchers to better understand the mobile usage and sensor data used in previous studies. Note that the data collected through the AS API (i.e., the main Android API in this study) can be further classified. For example, among the user interaction data corresponding to the fourth layer, the type of interaction data can be classified as click, long click, scroll, and typing once more. For the visible uniformity of the hierarchical structure, however, the three-layer hierarchical structure was mainly used in Section \ref{sec:datacategorization}. and user interactions were further explored in Section \ref{sec:userinteraction}.}

%% file: tables/existingdatacategory.tex
% \usepackage{booktabs}
% \usepackage{multirow}
% \usepackage{graphicx}

\begin{table}[]
\caption{Mobile usage and sensor data categories of surveyed studies}
\resizebox{\textwidth}{!}{%
\begin{tabular}{lll}
\toprule
\multicolumn{1}{c}{\textbf{Ref.}}   
& \multicolumn{1}{c}{\textbf{Category}} 
& \multicolumn{1}{c}{\textbf{Sub-Category}}                                                 

\\ \midrule\midrule
\multirow{2}{*}{Ferreira et al. ~\cite{ferreira2014contextual}}                             & "Application session"                   
& \begin{tabular}[c]{@{}l@{}}"when, for how long and which applications were active, provided by the Accessibility Service API screen usage"\end{tabular}                       

\\ \cmidrule(l){2-3} 
 & "Screen usage"                          
 & "current screen status (on/off) and for how long it was on/off"                                                                                                                  
 \\ \midrule
\multirow{3}{*}{\begin{tabular}[c]{@{}l@{}}Ferreira et al. ~\cite{ferreira2015aware}, \\ Church et al. ~\cite{church2015understanding}\end{tabular}}      
& "Hardware sensors"                      
& "Accelerometer, magnetometer, and photometer"                                                                                                                                         \\ \cmidrule(l){2-3} 
& "Software sensors"                      
& "User's calendar, email, social activity, and other logs (e.g., calls, messages)"                                                                                                     \\ \cmidrule(l){2-3} 
& "Human-based sensors"                   
& "mobile questionnaires (i.e., for ESM), voice, or gesture input"                                                                                                                      \\ \midrule
\multirow{2}{*}{Visuri et al. ~\cite{visuri2019understanding}}                              & "Contextual information"                
& "location, physical activity, headphone jack, ringer mode, screen state, battery information, network information, Foreground application"                                                                                                                            \\ \cmidrule(l){2-3} 
& "Notification information"              
& "source application contents notification outcome"                                        

\\ \midrule
\multirow{3}{*}{chang et al. ~\cite{chang2017smartphone}}                                   
& "Notification"                         
& \begin{tabular}[c]{@{}l@{}}"Sender application, arrival time, tag, ticker text, sound mode, vibrate mode, category, title, text, subtext, priority, visibility, \\ action, audio mode, clearable, ongoing of a notification-flag"\end{tabular}                                                                                                                           \\ \cmidrule(l){2-3} 
& "Context"                     
& \begin{tabular}[c]{@{}l@{}}"Lastest used, application and time, ringer mode, audio mode, volume, network and wifi status, location, physical activity, acceleration, \\ rotation, gravity, orientation, proximity, ambient light level, battery level, charging state"\end{tabular}                                                                          

\\ \cmidrule(l){2-3} 
& "Accessibility"                         
& "Triggered time, event type, event text, screen status"                                                                                                                                 \\ \midrule
\multirow{3}{*}{Lee et al. ~\cite{lee2014hooked}}                                           & "Application events"                    
& \begin{tabular}[c]{@{}l@{}}"Active/inactive apps, touch and text input events, web browsing URLs, and notification events"\end{tabular}                                                                                                                                    \\ \cmidrule(l){2-3} 
& "System events"                         
& "Power on/off and screen on/off/unlock"                                                                                                                                               \\ \cmidrule(l){2-3} 
& "Phone events"                          
& "Call and SMS"                                                                                                                                                                        \\ \midrule
\multirow{2}{*}{Pielot et al. ~\cite{pielot2017beyond}}                                     & "Phone state"                           
& "Ringer mode, battery level"                                                                                                                                                          \\ \cmidrule(l){2-3} 
& "Usage"
& "Screen use, app launches, and access to the notification center"                                                                                                                     \\ \midrule
\multirow{3}{*}{Zhang et al. ~\cite{zhang2017interaction}}                                  & "Event listeners"                       
& \begin{tabular}[c]{@{}l@{}}"Interface accessibility events (e.g., button click, text field focus, view update, app screen switch, device app switch)"\end{tabular}                                                                                                                 \\ \cmidrule(l){2-3} 
& "Content introspections"                
& "Element (e.g., content, size, state, possible actions)"                                                                                                                              \\ \cmidrule(l){2-3} 
& "Automation"                            
& \begin{tabular}[c]{@{}l@{}}"Accessibility service representation (e.g., click, long press,\\ select, scroll, or text input directed to a node in the tree)"\end{tabular}                                                                                                    \\ \midrule
\multirow{2}{*}{\begin{tabular}[c]{@{}l@{}}Schweizer et al. ~\cite{schweizer2014krakena}, \\ Schweizer et al. ~\cite{schweizer2014krakenb}\end{tabular}} & "Physical sensor"                       & \begin{tabular}[c]{@{}l@{}}"Sound pressure, ring mode, acceleration, activity, illuminance, network connectivity"\end{tabular}                                                                                                                                         \\ \cmidrule(l){2-3} 
& "Software sensor"                       
& "Browsing history, contents, calendar, location, call log, foreground app"                                                                                                            \\ \midrule
\multirow{7}{*}{Holzmann et al. ~\cite{holzmann2017android}}                                & "App usage data"                        
& \begin{tabular}[c]{@{}l@{}}"Visited pages ("Screens" within an app), dwell times (per page), number of interaction (per page), device orientation"\end{tabular}                                                                                                               \\ \cmidrule(l){2-3} 
& "Interaction data"                      
& "Type of interaction (touch, scroll, long touch), interaction target"                                                                                                                 \\ \cmidrule(l){2-3} 
& "Network data"                          
& "Network type, network subtype, roaming status"                                                                                                                                       \\ \cmidrule(l){2-3} 
& "Device data"                           
& \begin{tabular}[c]{@{}l@{}}"Device description, country of origin, language, network operator, operating system, screen resolution"\end{tabular}                                                                                                                             \\ \cmidrule(l){2-3} 
& "Battery data"                          
& "Battery level, charging status (charging/discharging), temperature, voltage"                                                                                                         \\ \cmidrule(l){2-3} 
& "Context data"                          
& "Light condition"                                                                                                                                                                     \\ \cmidrule(l){2-3} 
& "Orientation data"                      
& "Device orientation"                                                                                                                                                                  \\ \midrule
\multirow{2}{*}{Dingler et al. ~\cite{dingler2017language}}                                 & "Phone context"                         
& \begin{tabular}[c]{@{}l@{}}"Ringer mode (Mute, vibration, ringer), charging mode (unplugged, charging), battery status, display orientation (portrait or landscape mode, \\ orientation changes), light sensor, proximity sensor, location, motion"\end{tabular}                                                                                                    \\ \cmidrule(l){2-3} 
& "Phone usage"                           
& \begin{tabular}[c]{@{}l@{}}"Calls (incoming, outgoing), SMS (incoming, outgoing), notifications (received, dismissed, ignored, interacted with), screen (on/off events), \\ unlocks (phone unlocks), data usage (upload/download), applications (applications in foreground, switches, usage duration)"\end{tabular}                                                                                                                        
\\ \midrule
\multirow{3}{*}{Anderson et al. ~\cite{anderson2019impact}}                                 & "Notification"                          
& \begin{tabular}[c]{@{}l@{}}"Application (name, package), Notification time (issued, interacted, posted), Notification ongoing (true, false)"\end{tabular}                                                                                                                   \\ \cmidrule(l){2-3} 
& "Contacts"                              
& "Contact (hash), Relation (family, friend, work, none)"                                                                                                                              \\ \cmidrule(l){2-3} 
& "Events"                                
& \begin{tabular}[c]{@{}l@{}}"WiFi (SSID, BSSID), Location (latitude, longitude), Application (name, package), Time (date, issued, interacted, duration), ESM-Role (private, work), \\ ESM-Interruptibility (private, work, both, not interruptible), Physical Activity \\ (Google’s Activity Recognition API), Power (connected, disconnected), Screen (on, off), Ringer mode (silent, vibrate, normal)"\end{tabular} \\ 

\midrule
Chang et al. ~\cite{chang2015investigating}                                                 
& "Contexual information"                 
& \begin{tabular}[c]{@{}l@{}}"Location, activity recognition (Google activity recognition service API), sensors, network, calendar, phone status (ringer mode, screen on/off), \\ currently running application"\end{tabular}                                                                                                                                              \\ \midrule
\multirow{3}{*}{Chang et al. ~\cite{chang2017smartphone}}                                   & "Notification"                          
& \begin{tabular}[c]{@{}l@{}}"sender application, arrival time, tag, ticker text, sound mode, vibrate mode, category, title, text, subtext, priority, visibility, action, audio mode, \\ clearable, ongoing of a notification-flag"\end{tabular}                                                                                                                                  \\ \cmidrule(l){2-3} 
& "Context"                               
& \begin{tabular}[c]{@{}l@{}}"latest used, application and time, ringer mode, audio mode, volume, network and wifi status, location, physical activity, acceleration, rotation, \\ gravity, orientation, proximity, ambient light level, battery level, charging state"\end{tabular}                                                                         

\\ \cmidrule(l){2-3} 
& "Accessibility"                         
& "triggered time, event type, event text, screen status"                                                                                                                                                                           \\ \midrule
Lee et al. ~\cite{lee2019does}                               & "Contextual information"                
& "notifications, user's location and physical activity, activity type, the phone's status (e.g., ringer mode, battery level), sensor data, user actions"                              
\\ \bottomrule
\end{tabular}%
\label{table:existingdatacategorization}
}
\end{table}

%% file: 06results.tex
\section{Categorization Results of Collected Data and Research Purposes}
\label{sec:datacategorization}
This section illustrates data categorization results to help researchers to better understand the trend of data categories collected for various research purposes. In Section \ref{sec:resultsofdatacategorization}, overall results of data categorization are presented in Table \ref{table:datacategorization} which includes information on which themes, sub-themes, and IP types as depicted in Table \ref{table:IP} for each of 96 surveyed studies. In Section \ref{sec:resultsofdatacategorizationviaresearch}, data categories used for each theme are presented in Table \ref{table:datacategorization} and the analysis results are explained via Figure \ref{fig:numberofstudies}. In Section \ref{sec:resultsofdatacategorizationviasubresearch}, data categories used for each research theme and sub-theme are presented in Table \ref{table:researchtheme}. Furthermore, data types used in each theme and sub-theme are analyzed. \revision{This process provides the results} and interpretations of how the types of data collected are different according to the themes and sub-themes. Further, user interaction data are categorized and analyzed in Section \ref{sec:userinteraction}.

\input{tables/datacategorization}

\subsection{Collected Data Categorization}
\label{sec:resultsofdatacategorization}
This study analyzed the trends and differences \revision{in} data categories used and collected by each research theme in Table \ref{table:datacategorization}. The columns in Table \ref{table:datacategorization} are organized based on the second layer in Figure \ref{fig:datacategorization}. Only user interaction data are based on the third layer, because they are collected by using the main API (AS API) in this review. Among the UP (Usage Pattern) and NT (Notification) studies, 58.3\% (14 out of 24) and 78\% (14 out of 18) collected more than one data category in the system, context, and interaction sensing data category groups, respectively. 26.3\% (5 out of 19), 19.0\% (4 out of 21), 21.4\% (3 out of 14) surveyed studies belong to EUI/UX (Enhancement of UI and UX), PS (Privacy and Security), and PTS (Programming and Testing Support) themes collected and utilize system and context sensing data categories along with interaction sensing data categories. Through this, the researchers can infer that more varied types of mobile usage and sensor data are collected in UP and NT papers than EUI/UX, PS, and PTS studies.

\subsection{Research Purpose Specific Data Categorization}
\label{sec:resultsofdatacategorizationviaresearch}
\begin{figure*}
    \centering 
    \includegraphics[width=0.9\textwidth]{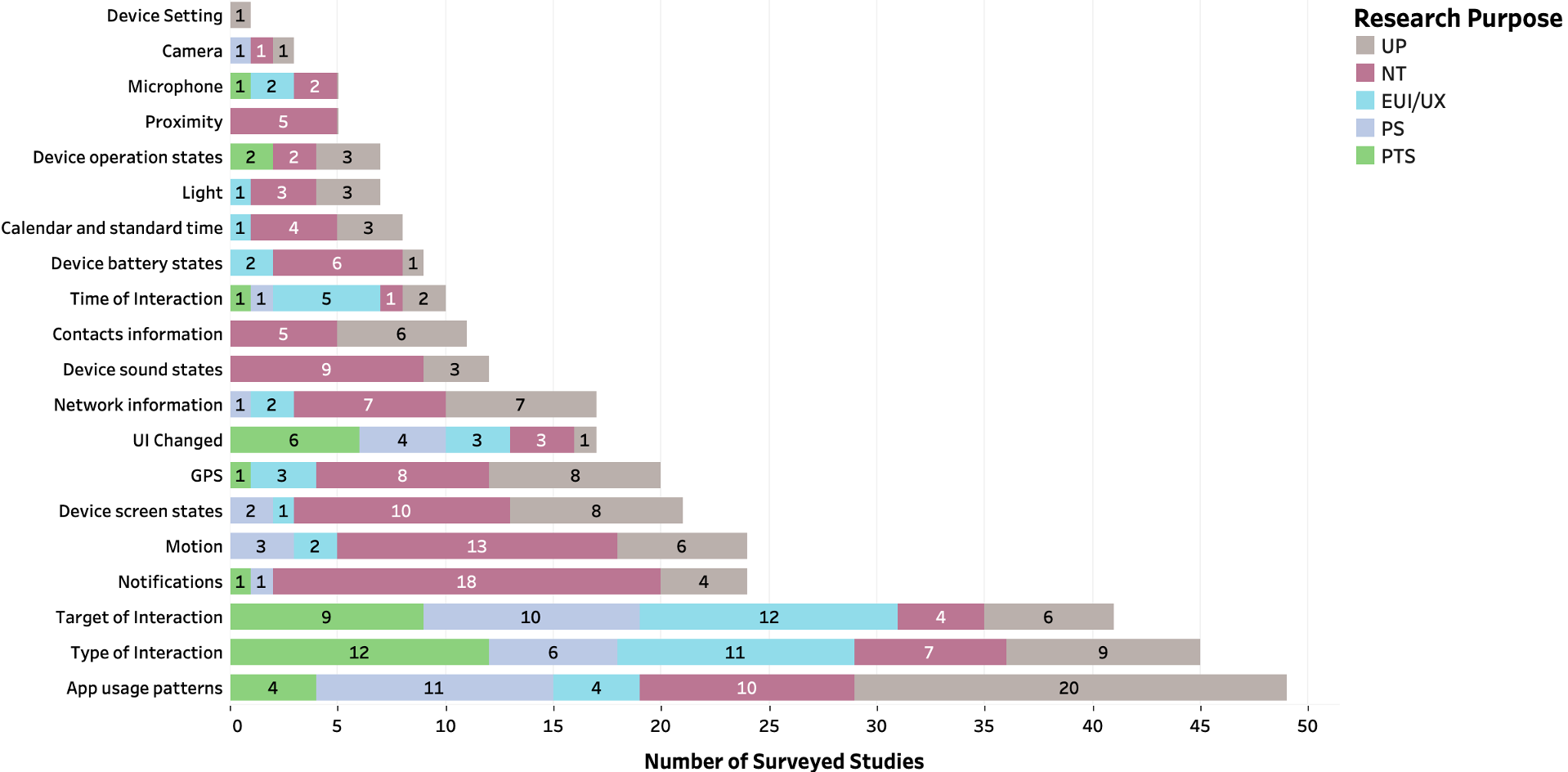} 
    \caption{Breakdown of research purposes for each mobile usage and sensor data item} 
    \label{fig:numberofstudies}
\end{figure*}

The number of studies that collected and used mobile usage and sensor data was identified for each research purpose as depicted in Figure \ref{fig:numberofstudies}. The frequently used data types were app usage patterns (49 studies), type of interaction (45 studies), the target of interaction (41 studies), motion (24 studies), notifications. (24 studies), device screen stats (21 studies), GPS (20 studies), UI changed (17 studies), and network information (17 studies).

Figure \ref{fig:numberofstudies} illustrates how data are used differently depending on the themes. In UP and NP themes, all data types were used in at least more than one surveyed study. In the UP theme, app usage pattern data was the most used of all data types, with 20 studies, and more than doubled the second most used data type, the interaction data type. Because of the 22 studies using the US API, nine of them were in the UP theme, and five of them only collected app usage pattern data. In addition, the device screen states, GPS, and network information data were widely used after app usage pattern data and type of interaction data with eight, eight, and seven, respectively. Accordingly, the researchers can infer that context and system sensing data were also widely used with interaction sensing data in the UP theme. In the NT's 18 surveyed studies, all data types except device setting data were used in more than one study. The most used data was notification data,used in 18 studies. Besides, motion, device screen status, app usage patterns, device sound states, GPS, types of interaction, network information data were frequently used. 

In contrast, most of the studies in EUI/UX, PS, and PTS themes mainly use interaction sensing data (i.e., user interaction, and app usage pattern). Among the 19 EUI/UX studies, target/type of interaction data was used in about 62\% of studies. App usage pattern and target of interaction data were used in approximately 50\% of 21 PS studies. In the PTS theme, target/type of interaction data was used in 12 (85.7\%) and 9 (64.29\%) of 14 studies.

\subsection{Sub-theme Level Categorization}
\label{sec:resultsofdatacategorizationviasubresearch}
This section reviewed how many surveyed studies were used in each data category for each sub-theme and analyzed the results with three measures as follows: the number of types of data categories, the sum of utilized data categories (i.e., calculated by counting each data category used in the papers), and the average number of data categories used per surveyed study (i.e., the sum of utilized data categories / total number of surveyed studies) by each sub-theme as depicted in Table \ref{table:researchtheme}. To easily understand the number of data categories used by each sub-theme in Table \ref{table:researchtheme}, cells were marked with a darker green color as the number of papers increased. Moreover, Table \ref{table:researchtheme} is displayed with dark red cells as the number increases to easily distinguish the number of papers for each sub-theme and the trend calculated by three measures. Accordingly, this study synthesized the results of Section \ref{sec:resultsofdatacategorization} and \ref{sec:resultsofdatacategorizationviaresearch}, and interpreted the trends in utilization of data categorization for each research theme/ sub-theme.

\subsubsection{Usage Pattern (UP) Research Theme} The number of data categories used of FSD (Framework and System Development) sub-theme studies \revision{were} higher compared with other sub-themes studies in the UP theme as described in Table \ref{table:researchtheme}. Most studies of FSD develop third-party apps to collect various types of mobile usage and sensor data. By contrast, the studies of GUPA (Generic Usage Pattern Analysis) and MLMS (Machine Learning-based Model and System) sub-themes mainly used interaction sensing data. Accordingly, the average used data categories was \revision{2.6,} which is lower compared to FSD (6.0) as shown in Table \ref{table:researchtheme}. When the AS API was used among surveyed studies of GUPA and MLMS sub-themes, various mobile usage and sensor data were collected by using other Android APIs as well. However, using the US API, the average of data categories used per surveyed study was relatively low. It only needs app usage patterns about the current usage status of the device system or third-party apps rather than detailed contextual information. 

\subsubsection{Notification (NT) Research Theme} NT theme studies collected and utilized various data categories compared to other themes. As shown in Table \ref{table:researchtheme}, the NT studies had higher values in three measures  compared to other themes (i.e., the sum of utilized data categories, number of types of data categories, and average number of data categories used per surveyed study). In addition, the proportion of using context sensing and system sensing data along with interaction sensing data was the highest compared to other themes according to the analysis in Section \ref{sec:resultsofdatacategorizationviasubresearch}. The reasons why NT theme studies have utilized various data categories are as follows. Studies of IFN (Identifying Factors respond to Notifications) sub-theme utilized a variety of sensor data to predict in which context a specific notification interacts with the user~\cite{visuri2019understanding, pielot2014didn, chang2015investigating}. Studies of HNAU (How Notification Affects a User) sub-theme collected user interaction data to understand the effects of notifications according to various situations on users~\cite{pielot2017beyond, pradhan2017understanding, park2017don}. In addition, studies of NT/NM (Notification Timing and Notification Management) sub-theme collected various types of mobile data to find the right notification timing and manage notification in various contexts~\cite{chang2017smartphone, komuro2017relationship, anderson2019impact,lee2019does, dingler2017language}.

\input{tables/researchtheme}

\subsubsection{Enhancement of UI/UX (EUI/UX) Research Themes} Studies of EUI/UX theme collect and utilize on average 2.72 data types per a study as shown in Table \ref{table:researchtheme}. Since most studies in the EUI/UX theme were conducted to improve the user interface, only user interaction data were collected and analyzed by using the AS API. Only five studies collected context sensing data types along with interaction sensing data types~\cite{sun2018lip, tarakji2018voice, rauen2018empowering, lin2019user, chen2018construction}.

\subsubsection{Privacy and Security (PS) Research Themes} As shown in Table 10, the average number of data categories used per the surveyed study of PS theme was 2.05 which means most studies of PS theme used only a few interaction data types. The studies of APC (Access and Permission Control), PPS (Privacy-Preserving System), and MDPS (Monitoring and Detection for Privacy and Security), which are sub-themes of PS, mainly focused on user interaction data collection and utilization. Among them, a total of 15 studies related to privacy and security collected and utilized user interaction data types such as user interaction type/target or UI changed data types~\cite{bonne2017exploring, rahman2018iac, fawaz2016privacy, lau2014mimesis, ozcan2015babelcrypt, andriotis2020allow, fernandes2016appstract, naseri2019accessileaks, rastogi2016these, shao2018understanding, vishwamitra2017mcdefender, leguesse2020reducing, li2018chatting, wang2019dcdroid, wang2020identifying}. In contrast, studies of \revision{the} ASS (Authentication System/Scheme) sub-theme used the context and system sensing data types along with user interaction data types to understand users' daily habits and current device hold situations or develop the motion sensor-based authentication system~\cite{aras2019multilock, kraus2017use, canfora2016silent, zhu2019riskcog, zhu2020hybrid, torres2019behavioral}.

\subsubsection{Programming and Testing Support (PTS) Research Themes} Most of the studies related to the PTS theme collected and utilized only a few user interaction data categories (the average number of data categories per study was 2.62). In addition to these data, two studies have collected device operating state data for system performance measurement and management~\cite{bissig2018towards, xiang2020dynamical, li2017sugilite}. \revision{Except for} one study using microphone data~\cite{li2017sugilite}, studies of GAT (GUI Automated Testing), CT (Crowdsourced Testing), APS (Automated Programming Support), and TRSGT (Test Recorder and Script Generator Technique) sub-themes mainly used AS API-based interaction sensing data types for crowdsourced testing, GUI automated testing, user interaction recording, automated programming support~\cite{arruda2016capture, gu2019practical, muangsiri2017random, lian2018cat, wu2017appcheck, fazzini2017barista, liu2017mechanism, negara2019practical, li2017programming, li2020interactive, li2018kite, li2017sugilite}.

\subsection{User Interaction Level Categorization}
\label{sec:userinteraction}
The type of touch interaction data (e.g., click, long click, typing, scroll) were further classified and analyzed for each surveyed study. Among a total of 44 such studies, 18 studies used terms related to the types of touch interaction actions (i.e., click, long click, typing, scroll, pinch, swipe), and the 15 studies used the event name of \emph{AccessibilityEvent} such as \revision{VIEW\_CLICKED, VIEW\_LONG\_CLICKED, VIEW\_TEXT\_CHANGED, VIEW\_SCOLLED} to describe the collection of user’s touch interaction data. The remaining 11 studies used general terms (e.g., user action, touch event, all UI events, touch event, input action, gesture action) in their papers. As shown in Table \ref{table:touchinteractions}, these studies were grouped and classified into five categories: general expression, click (including tapping), long click, typing, and scroll (including punch and swipe). \revision{Click and typing were most frequently used in previous research because they are the most basic interactions in smartphone use.}
\input{tables/touchinteraction}
Among the studies using AS API, 17 studies described \emph{AccessibilityEvent} terms in their papers. As shown in Table \ref{table:eventtype}, \emph{AccessibilityEvent} was divided into four types: view types (e.g., \revision{VIEW\_CLICKED}), transition types (e.g., \revision{WINDOW\_STATE\_CHANGED}), notification types (e.g., \revision{NOTIFICATION\_STATE\_CHANGED}), and exploration types (e.g., \revision{TOUCH\_INTERACTION\_START}). The sum of accessibility events belonging to the four types \emph{AccessibilityEvent} described in these 17 studies are 53, 21, 2, and 2, respectively. Accordingly, the view and transition types are described more than the notification and exploration types in the 18 studies. Although notification data was used in most NP theme studies, only two papers described notification types of \emph{AccessibilityEvent} (NOTIFICATION\_STATE\_CHANGED). \revision{This tendency shows that notification types of AccessibilityEvent were not properly described in the previous papers.} In addition, the exploration type was described less frequently because of its low utility rather than described in a different way in the paper. Among the \emph{AccessibilityEvent}, the most frequent view type events were VIEW\_CLICKED.
\input{tables/eventtype}
In addition to 18 studies describing \emph{AccessibilityEvent} used in Table \ref{table:eventtype}, we deduced which \emph{AccessibilityService} was used for 24 surveyed studies that used the AS API-based interaction sensing data but did not describe which types of \emph{AccessibilityEvent} were used to detect the interaction sensing data. According to \emph{AccessibilityEvent} that are inferred to have been used to capture the utilized data, the list of surveyed studies is as follows.

\begin{itemize}
\item \textbf{Notification types (i.e., \revision{NOTIFICATION\_STATE\_CHANGED})}: 18 studies used notification data~\cite{blanke2014mining, lee2014hooked, andone2016menthal, pielot2014didn, visuri2019understanding, chang2015investigating, lee2018exploring, okoshi2015attelia, okoshi2016towards, park2017don, pielot2017beyond, anderson2019impact, dingler2017language, komuro2017relationship, lee2019does, pielot2014situ, fawaz2016privacy, li2017programming}
\item \textbf{Exploration types (i.e., TOUCH\_GESTURE\_DETECTION\_START/END)}: \revision{Two studies used gesture interaction data~\cite{rodrigues2015breaking, alshayban2020accessibility}}
\item \textbf{View types (i.e., \revision{VIEW\_CLICKED})}: 42 studies in Table \ref{table:touchinteractions} (except~\cite{alshayban2020accessibility, rodrigues2015breaking, mehrotra2017understanding})
\item \textbf{Transition types (i.e., \revision{WINDOW\_STATE\_CHANGED})}: 18 studies used UI changed data~\cite{lee2018click, okoshi2015attelia, okoshi2016towards, chang2017smartphone, lin2019user, riegler2018measuring, alshayban2020accessibility, rahman2018iac, lau2014mimesis, fernandes2016appstract, li2018chatting, muangsiri2017random, lian2018cat, li2020interactive, li2017sugilite, bissig2018towards, fazzini2017barista} and 30 studies that used app usage pattern data~\cite{blanke2014mining, church2015understanding, kim2019understanding, lee2014hooked, ferreira2014contextual, lee2018click, andone2016menthal, ferreira2015aware, holzmann2017android, schweizer2014krakena, schweizer2014krakenb, visuri2019understanding, chang2015investigating, pielot2017beyond, pradhan2017understanding, anderson2019impact, chang2017smartphone, dingler2017language, li2018appinite, rauen2018empowering, rodrigues2015getting, alshayban2020accessibility, aras2019multilock, kraus2017use, lau2014mimesis, li2018chatting, li2017programming, bissig2018towards, fazzini2017barista}
\end{itemize}

%% file: tables/datacategorization.tex
\begingroup % localize the following settings                                                       

\setlength\LTcapwidth{\textwidth} % default: 4in (rather less than \textwidth...)                                                                                                                           
\setlength\LTleft{0pt}            % default: \parindent                                                                                                                                                     
\setlength\LTright{0pt}           % default: \fill                                              
\tiny
\setlength{\tabcolsep}{2.7pt}
%\begin{adjustbox}{max width=\textwidth}
%
%\end{adjustbox}

\endgroup

%% file: tables/researchtheme.tex
\begin{table}
\caption{Number of studies that used based on the types of data in each research and sub-research theme}
\label{table:researchtheme}
\centering
\begin{adjustbox}{max width=\textwidth}
\begin{tabular}{cc ccccccc c cccccc c ccccccc c cccc}
\toprule

\multirow{14}{*}{\textbf{\shortstack{Theme}}} & 
\multirow{14}{*}{\textbf{\shortstack{Sub-Theme}}} & 
\multicolumn{7}{c}{\textbf{Interaction sensing}} & 
\multirow{14}{*}{} &
\multicolumn{6}{c}{\textbf{System sensing}} &
\multirow{14}{*}{} &
\multicolumn{7}{c}{\textbf{Context sensing}} & 
\multirow{14}{*}{} &
\multirow{14}{*}{\raisebox{-12em}{\rotatebox{90}{\textbf{\shortstack{\# of types of data categories}}}}} & 
\multirow{14}{*}{\raisebox{0.5em}{\rotatebox{90}{\textbf{\shortstack{Sum of utilized data categories}}}}} & 
\multirow{14}{*}{\raisebox{-12em}{\rotatebox{90}{\textbf{\shortstack{\# of surveyed studies}}}}} & 
\multirow{14}{*}{\raisebox{-12em}{\rotatebox{90}{\textbf{\shortstack{Average of data categories\\ \hspace{-0.25em}used per surveyed study}}}}} \\
\bottomrule
\end{tabular}
\end{adjustbox}
\end{table}

%% file: tables/touchinteraction.tex



\begin{table}[]
\caption{Collected touch interaction type information from the surveyed papers}
\resizebox{\textwidth}{!}{%
\centering 
\begin{tabular}{lm{13cm}r}
\toprule
\multicolumn{1}{c}{\textbf{Types of Touch Interaction}} & \multicolumn{1}{c}{\textbf{Ref.}} & \multicolumn{1}{c}{\textbf{Num}} \\ \midrule\midrule
General Expression & ~\cite{rahman2018iac, lee2019does, chang2017smartphone, ferreira2015aware, park2017don, sun2018lip, rauen2018empowering, li2017programming, alshayban2020accessibility, rodrigues2015breaking} & 10 \\ \hline
Click (including tapping) & ~\cite{kim2019understanding, lee2018click, holzmann2017android, montague2015tinyblackbox, okoshi2015attelia, okoshi2016towards, elazhary2018w, tarakji2018voice, lin2019user, riegler2015ui, riegler2018measuring, fernandes2016appstract, arruda2016capture, gu2019practical, muangsiri2017random, lian2018cat, wu2017appcheck, li2018kite, li2017sugilite, bissig2018towards,  fazzini2017barista, liu2017mechanism, negara2019practical, zhong2015enhancing, mehrotra2017understanding} & 25 \\ \hline
Long click & ~\cite{kim2019understanding, lee2018click, holzmann2017android, okoshi2015attelia, okoshi2016towards, riegler2018measuring, muangsiri2017random, lian2018cat, wu2017appcheck, li2018kite, li2017sugilite, negara2019practical, mehrotra2017understanding} & 13 \\ \hline
Typing & ~\cite{lee2014hooked, bitsch2015psychologist, schweizer2014krakena, schweizer2014krakenb, okoshi2015attelia, okoshi2016towards, elazhary2018w, tarakji2018voice, chen2019messageontap, lin2019user, canfora2016silent, naseri2019accessileaks, li2018chatting, muangsiri2017random, lian2018cat, wu2017appcheck, li2018kite, li2017sugilite, fazzini2017barista, negara2019practical}  & 20 \\ \hline
Scroll (Including Pinch and Swipe) & ~\cite{kim2019understanding, holzmann2017android, okoshi2015attelia, okoshi2016towards, tarakji2018voice, lin2019user, riegler2015ui, riegler2018measuring, lau2014mimesis, li2018chatting, gu2019practical, muangsiri2017random, lian2018cat, wu2017appcheck, zhong2015enhancing, mehrotra2017understanding, negara2019practical, montague2015tinyblackbox} & 18 \\ \bottomrule
\end{tabular}%
\label{table:touchinteractions}
}
\end{table}

%% file: tables/eventtype.tex
\begin{table}
\caption{Accessibility events described in surveyed studies}
\centering
\begin{adjustbox}{max width=\textwidth}
\begin{tabular}{lccc@{\hskip 0.3in}cccccccc@{\hskip 0.3in}cccccc@{\hskip 0.3in}ccc@{\hskip 0.3in}ccrc}
%\begin{tabular}{llllllllllllllllllllllll}
\toprule

\multirow{10}{*}{\raisebox{-1em}{\textbf{Ref.}}} & 
\multirow{10}{*}{\raisebox{-1em}{\textbf{Theme}}} & 
\multirow{10}{*}{\raisebox{-1em}{\textbf{Sub-Theme}}} &
\multirow{10}{*}{\raisebox{-3em}{\textbf{\begin{tabular}[c]{@{}c@{}}Type of\\Included\\Papers\end{tabular}}}} &
\multicolumn{8}{m{4cm}}{\textbf{VIEW TYPES}} & \multicolumn{6}{l}{\textbf{TRANSITION TYPES}} & \multicolumn{3}{c}{\textbf{NOTIFICATION TYPES}} & \multicolumn{4}{c}{\textbf{EXPLORATION TYPES}} \\
\cmidrule{5-25}

& & & & 

\multirow[b]{9}{*}{\raisebox{-0.9em}{\rotatebox{90}{\textbf{CLICKED}}}} &
\multirow[b]{9}{*}{\raisebox{-0.9em}{\rotatebox{90}{\textbf{LONG CLICKED}}}} &
\multirow[b]{9}{*}{\raisebox{-0.9em}{\rotatebox{90}{\textbf{SELECTED}}}} &
\multirow[b]{9}{*}{\raisebox{-0.9em}{\rotatebox{90}{\textbf{FOCUSED}}}} &
\multirow[b]{9}{*}{\raisebox{-0.9em}{\rotatebox{90}{\textbf{TEXT CHANGED}}}} &
\multirow[b]{9}{*}{\raisebox{-0.9em}{\rotatebox{90}{\textbf{\shortstack{TEXT TRAVERSED\\\hspace{-1.4em} AT MOVEMENT\\\hspace{-1.3em}GRANULARITY}}}}} &
\multirow[b]{9}{*}{\raisebox{-0.9em}{\rotatebox{90}{\textbf{\shortstack{TEXT SELECTION\\\hspace{-3.2em}CHANGED}}}}} &
\multirow[b]{9}{*}{\raisebox{-0.9em}{\rotatebox{90}{\textbf{SCROLLED}}}} &

\multirow[b]{9}{*}{\raisebox{-0.9em}{\rotatebox{90}{}}} &
\multirow[b]{9}{*}{\raisebox{-0.9em}{\rotatebox{90}{}}} &
\multirow[b]{9}{*}{\raisebox{-0.9em}{\rotatebox{90}{\textbf{\shortstack{WINDOW STATE\\\hspace{-2.8em}CHANGED}}}}} &
\multirow[b]{9}{*}{\raisebox{-0.9em}{\rotatebox{90}{\textbf{\shortstack{WINDOW CONTENT\\\hspace{-4.7em}CHANGED}}}}} &
\multirow[b]{9}{*}{\raisebox{-0.9em}{\rotatebox{90}{\textbf{WINDOW CHANGED}}}} &
\multirow[b]{9}{*}{\raisebox{-0.9em}{\rotatebox{90}{}}} &

\multirow[b]{9}{*}{\raisebox{-0.9em}{\rotatebox{90}{}}} &
\multirow[b]{9}{*}{\raisebox{-0.9em}{\rotatebox{90}{}}} &
\multirow[c]{9}{*}{\raisebox{-0.9em}{\textbf{\shortstack{NOTIFICATION STATE\\CHANGED}}}} &

\multirow[b]{9}{*}{\raisebox{-0.9em}{\rotatebox{90}{}}} &
\multirow[b]{9}{*}{\raisebox{-0.9em}{\rotatebox{90}{}}} &
\multirow[b]{9}{*}{\raisebox{-0.9em}{\rotatebox{90}{\textbf{\shortstack{\hspace{-4.4em}TYPE TOUCH\\INTERACTION START}}}}} &
\multirow[b]{9}{*}{\raisebox{-0.9em}{\rotatebox{90}{\textbf{\shortstack{\hspace{-3.3em}TYPE TOUCH\\INTERACTION END}}}}} \\

& & & & & & & &\\
& & & & & & & &\\
& & & & & & & &\\
& & & & & & & &\\
& & & & & & & &\\
& & & & & & & &\\
& & & & & & & &\\
& & & & & & & &\\
& & & & & & & &\\

\midrule\midrule

Lee et al. ~\cite{lee2018click} & UP & MLMS & IP1-1-AS & 
$\bullet{}$ & $\bullet{}$ &  &  &  &  &  &  &  &  & $\bullet{}$ & &  &             &             &             &  &             &             &  & \\
 
Okoshi et al. ~\cite{okoshi2015attelia} & NT & NT/NM & IP1-1-AS & 
$\bullet{}$ & $\bullet{}$ & $\bullet{}$ & $\bullet{}$ & $\bullet{}$ & $\bullet{}$ & $\bullet{}$ &
$\bullet{}$ &             &             & $\bullet{}$ & $\bullet{}$ &  &             &             &             & $\bullet{}$ &             &             &  & \\
 
Okoshi et al. ~\cite{okoshi2016towards} & NT & NT/NM & IP1-1-AS & 
$\bullet{}$ & $\bullet{}$ & $\bullet{}$ & $\bullet{}$ & $\bullet{}$ & $\bullet{}$ & $\bullet{}$ &
$\bullet{}$ &             &             & $\bullet{}$ & $\bullet{}$ & &             &             &             & $\bullet{}$ &             &             &  & \\

Riegler et al. ~\cite{riegler2018measuring} & EUI/UX & UIP & IP1-1-AS & 
$\bullet{}$ & $\bullet{}$ & $\bullet{}$ & $\bullet{}$ &  &  &  &
$\bullet{}$ &             &             & $\bullet{}$ & $\bullet{}$ & $\bullet{}$ &             &             &             &  &             &             & $\bullet{}$ & $\bullet{}$\\

Rodrigues et al. ~\cite{rodrigues2018aidme} & EUI/UX & IFASD & IP2-1 & 
 &  & $\bullet{}$ & $\bullet{}$ &  & & &
 &             &             &  &  &  &             &             &             &  &             &             &  & \\
 
Aras et al. ~\cite{aras2019multilock} & PS & ASS & IP1-1-AS & 
 &  & &  &  &  &  & &             &             &  & $\bullet{}$ &  &             &             &             &  &             &             &  & \\

Lau et al. ~\cite{lau2014mimesis} & PS & PPS & IP1-1-AS & 
&  &  & &  & &  &
$\bullet{}$ &             &             & $\bullet{}$ & $\bullet{}$ &  &             &             &             & &             &             &  & \\

Fernandes et al. ~\cite{fernandes2016appstract} & PS & PPS & IP1-2-AS & 
$\bullet{}$ & & $\bullet{}$ & &  & &  & &             &             &  & $\bullet{}$ & &             &             &             & &             &             &  & \\

Naseri et al. ~\cite{naseri2019accessileaks} & PS & MDPS & IP1-1-AS & 
& & & & $\bullet{}$ & $\bullet{}$ &  & &             &             &  &  &  &             &             &             &  &             &             & & \\

Li et al. ~\cite{li2018chatting} & PS & MDPS & IP1-2-AS & 
 & &  &  &  & & &
$\bullet{}$ &             &             &  & $\bullet{}$ &  &             &             &             & &             &             &  & \\

Muangsiri et al. ~\cite{muangsiri2017random} & PTS & GAT & IP1-2-AS & 
$\bullet{}$ & $\bullet{}$ & & & $\bullet{}$ &  & &
$\bullet{}$ &             &             & $\bullet{}$ & &  &             &             &             &  &             &             &  & \\

Lian et al. ~\cite{lian2018cat} & PTS & CT & IP1-2-AS & 
$\bullet{}$ & $\bullet{}$ & $\bullet{}$ & & & $\bullet{}$ &  &
$\bullet{}$ &             &             &  &  & $\bullet{}$ &             &             &             & &             &             & & \\

Li et al. ~\cite{li2018kite} & PTS & APS & IP1-2-AS & 
$\bullet{}$ & $\bullet{}$ &  &  & $\bullet{}$ &  &  &
 &             &             & &  &  &             &             &             &  &             &             &  & \\

Li et al. ~\cite{li2017sugilite} & PTS & APS & IP1-2-AS & 
$\bullet{}$ & $\bullet{}$ & $\bullet{}$ & $\bullet{}$ & $\bullet{}$ & &  &
 &             &             & $\bullet{}$ & $\bullet{}$ &  &             &             &             &  &             &             &  & \\

Bissig et al. ~\cite{bissig2018towards} & PTS & PMM & IP1-1-AS & 
$\bullet{}$ & &  &  &  &  &  &
 &             &  & $\bullet{}$ & $\bullet{}$ & &             &             &             & &             &             & & \\

Fazzini et al. ~\cite{fazzini2017barista} & PTS & TRSGT & IP1-1-AS & 
$\bullet{}$ &  &  &  & $\bullet{}$ &  &  &
&             &             & $\bullet{}$ & $\bullet{}$&  &             &             &             &  &             &             &  & \\

Liu et al. ~\cite{liu2017mechanism} & PTS & TRSGT & IP1-2-AS & 
$\bullet{}$ &  & & $\bullet{}$ & & &  &
 &             &             &  &  &  &             &             &             &  &             &             & & \\

\bottomrule
\end{tabular}
\end{adjustbox}
\label{table:eventtype}
\end{table}

%% file: 07discussion.tex
\section{Discussion}
\label{sec:discussion}
Several issues and implications found while investigating existing studies are discussed in this section. \revision{In Section \ref{sec:summaryandconsideration}, a summary of categorization results is presented.} \revision{In Section \ref{sec:description}, the difficult cases to identify the smartphone data terms are summarized.} \revision{In Section \ref{sec:APIhistory}, the impact of changes via the API version on research is reported, and the differences due to changes in API policy are discussed.} In Section \ref{sec:privacy}, privacy issues that may arise in terms of data collection in mobile usage and sensor data-driven analytics research are analyzed. \revision{In Section \ref{sec:quality}, factors and solutions to data quality degradation occurring in mobile usage and sensor data-driven research are briefly reviewed.} In Section \ref{sec:limitation}, limitations and future research through this review are discussed.

\subsection{\revision{Summary of Categorization Results}}
\label{sec:summaryandconsideration}

This categorization results in this work helped researchers to understand what data was used for what research purpose. First, this research identified which data categories were used for each surveyed study via the research theme, sub-theme, and type of included papers (IP), as shown in Table \ref{table:datacategorization}. However, it was difficult to grasp at a glance what data was used for each research purpose through Table \ref{table:datacategorization}. Accordingly, this work analyzed which data categories were used for each research theme, as shown in Figure \ref{fig:numberofstudies}. The results showed that the trends of the used data types are different according to each research theme. Respectively, 83\% and 58\% of the NT and UP studies utilized context and system sensing data as well as interaction sensing data. Unlike NT and UP studies, prior EUI/UX, PS, and PTS studies mainly used interaction sensing data (74\%, 81\%, and 80\%, respectively), and thus, context and system sensing data usage was relatively low.

Further, the data types used by sub-themes classified in each research theme were analyzed to understand the data types used for each research purpose in more detail, as shown in Table \ref{table:researchtheme}. As a result, different trends were shown for the data types used according to the sub-themes for each research theme. In NT and UP studies, data types corresponding to context sensing or system sensing data in all sub-themes except one sub-theme (i.e., MLMS) were utilized along with data types belonging to interaction sensing data. In the EUI/UX studies, there has been at least one research using data types belonging to context sensing or system sensing data in addition to interaction sensing data in four sub-themes (i.e., CE, UIE, UIP, DEASD) out of six sub-themes. In the PS and PTS studies, six sub-themes (i.e., APC, PPS, MDPS, GAT, CT, TRSGT) out of nine sub-themes only used interaction sensing data. Our analysis results helped researchers and practitioners identify the tendency of data types utilized according to various research purposes and derive new insights for selecting the range of data items to be collected for research planning.

\subsection{Reproducibility Risks Due to a Lack of Standardized Data Typology}
\label{sec:description}
While investigating the mobile usage and sensor data collected and used in the studies, the identification of mobile usage and sensor data terms in the studies can be divided into (1) the detailed explanation of data terms and (2) the vague description of data terms. As shown in Table \ref{table:existingdatacategorization}, for studies of (1) in which it is easy to identify the used events and data terms are clear, and the data are categorized with the description of each data and displayed in a table. In contrast, for studies of (2) in which the API and data names are not explicitly described, or the scope of the data is unclear because only a few examples are described. For example, it is difficult to identify what data are collected by specific papers that used vague terms (e.g., "UI Interaction Collection," "sensors," and "sensor data'') rather than specifying what UI interaction data or sensor data are utilized using the AS API. Next, there are some papers where it is difficult to understand the entire data used as these papers only described a few examples using “such as,” “e.g.'' and “ex.” The specification of the collected data described in case (2) papers do not interfere with the understanding of the papers. However, if the data specification is described more specifically as in case (1), the researcher's understanding of the data can be much clearer. In that case, it can be communicated more clearly to the readers to identify the data which are described in each paper. The studies were classified into two types according to the degree of the detailed description in the papers in terms of data categorization based on the hierarchical structure:

\begin{itemize}
\item\textbf{{Detailed described studies as the data term level of the third layer in Figure \ref{fig:numberofstudies} (e.g., "click," "scroll," "text changed," "notification time," "accelerometer," "gyro," and "battery level")}}~\cite{visuri2019understanding, pielot2014didn, anderson2019impact, komuro2017relationship, pradhan2017understanding, pielot2014situ, church2015understanding, lee2014hooked, kim2019understanding, okoshi2015attelia, okoshi2016towards, lee2018exploring, andone2016menthal, holzmann2017android, schweizer2014krakena, schweizer2014krakenb, lee2018click, blanke2014mining, dingler2017language, ferreira2014contextual, lin2019user, riegler2018measuring, arruda2016capture, bissig2018towards, fazzini2017barista, negara2019practical, lau2014mimesis, naseri2019accessileaks}
\item\textbf{{Vaguely described studies as the data term level of the first or second layer in Figure \ref{fig:numberofstudies} (e.g., "all UI events," "sensors," "sensor data," "event type," "user actions," "gesture action," "context events," "context logs")}}~\cite{lee2019does, chang2015investigating, chang2017smartphone, rodrigues2015breaking, park2017don, zheng2017cleaning,  wang2018client, san2016gui}
\end{itemize}

\begin{figure*}
    \centering 
    \includegraphics[width= \textwidth]{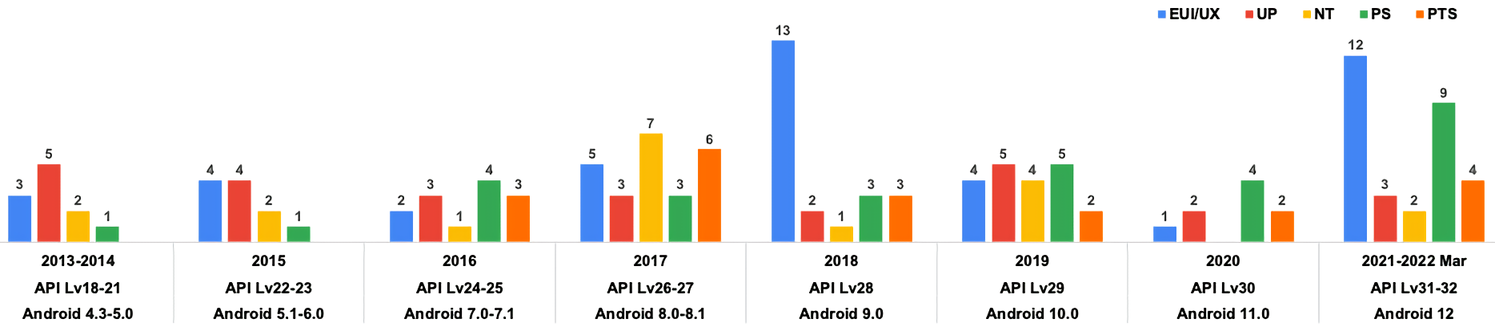} 
    \caption{Impact of Research according to API Changes History} 
    \label{fig:historyAPI}
\end{figure*}

Furthermore, depending on the method of describing the data terms, there can be a difference in the degree of ease in terms of reproducibility for other researchers. The way of representing data terms in 110 papers can be divided into two types of studies: The first type of studies are described by words or words with categorization, and words in sentences describe the second type of studies. As shown in Table \ref{table:existingdatacategorization}, the data term descriptions in the first type of study are more straightforward to understand than those of the second type of study.  Besides, there are surveyed studies that described the Android-related component terms (e.g., API, class, event, smartphone data) as described in the Android or Google developer documentation. In addition, there are surveyed studies that did not refer to the terms in the Android or Google developer documentation. Researchers who are new to data-driven research, instead of experienced researchers who have conducted smartphone-related research, would find it easier to grasp and reproduce papers with data terms written according to Android or Google developer documentation. Therefore, if the terms used to describe events and data match with that in the Google developer documentation, future researchers can understand the related research more quickly in this field.

\subsection{Influence of API Updates on Research}
\label{sec:APIhistory}
\revision{This subsection reports the influence of Android APIs (e.g., AS/US API, NotificationListenerService, NotificationManager) release and update on the research trend. As shown in Figure \ref{fig:historyAPI}, this study derived the number of reviewed papers for each research theme via API level, year, and Android version. The releases and updates of AS/US API from 2008 to 2022 are described in detail in Appendix D Table A5. During 2008--2012, NotificationManager (API level 1) and AS API (API level 4) were published and updated (API level 9--17)~\cite{NotificationManager, AccessibilityEvent, AccessibilityService}.} \revision{Major events of AS API were mostly updated which are view types (e.g., view clicked/long clicked/selected/focused/text changed/scroll/text selection changed), transition types (e.g., window state/content changed), exploration types (e.g., view hover enter/exit, touch interaction, touch exploration gesture), and type notification state changed at the API level 4--17~\cite{AccessibilityEvent}}.  
%\subsubsection{2008–2012: AS API \& Notification API release}

%AS/US and Notification API release and update were analyzed to find their influence on the research trend. As shown in Figure \ref{fig:historyAPI}, NotificationManager (API level 1) and AS API (API level 4) were published during 2008–2012 (API level 1–17) and were continuously updated in API 9–17~\cite{NotificationManager, AccessibilityEvent, AccessibilityService, NotificationListenerService}. When the AS API was first provided to develop third-party apps in API level 4, most of \emph{AccessibilityEvent} was published to track the type of interaction, notification, and window state changed~\cite{AccessibilityEvent}. Since API level 14–17 in 2011–2013, the touch exploration mode has been added in AS API. Furthermore, \emph{AccessibilityEvent} such as window content changed, view scroll, and view text selection changed has been added from Android 4.0 (API level 14). According to the AS API function update, the results of the study, which started in late 2011–early 2012, were first published on July 31, 2013, under LG Sangam Library's digital accessible information system for the disabled~\cite{hann2013best}. 

\emph{\revision{2013--2014: US API and NotificationListenerService API Release \& Start Publishing Research using AS API}}: \revision{According to the AS API function update in API level 1--17~\cite{NotificationManager, AccessibilityEvent, AccessibilityService, NotificationListenerService}, the results of the study, which started in late 2011--early 2012, were first published on July 31, 2013, under LG Sangam Library's digital accessible information system for the disabled~\cite{hann2013best}.} In 2014, a total of seven studies of the theme (UP: 5, EUI/UX: 2, NT: 2, PS: 1) were published in earnest based on the functions updated through API 14--17 from 2011 to 2013 as shown in Figure \ref{fig:historyAPI}. \revision{In API 14--17, Through touch exploration mode which provides feedback (e.g., voice, vibration) on touch content by detecting \emph{AccessibilityEvent}~\cite{AccessibilityEvent} generated from user's gesture/touch interaction (e.g., gesture detection start/end, touch navigation, view hover enter), studies for the visually impaired corresponding to DEASD and IFASD sub-theme were published~\cite{zhong2014justspeak, ganz2014percept}.} Further, studies using AS API to monitor notification states changed and app usage status were published for the first time in NP theme~\cite{pielot2014didn, pielot2014situ}. Moreover, the UP theme has three studies of GUPA sub-theme that use AS API to understand the time, type, and frequency of app usage such as smartphone overuse~\cite{blanke2014mining, ferreira2014contextual, lee2014hooked}, and two studies under the FSD sub-theme \revision{were} published that developed a framework for collecting app usage logs and ambient contextual data using smartphone sensors using SensorManager (released in API level 3) and AS API~\cite{schweizer2014krakena, schweizer2014krakenb}. Finally, the purpose of AS API-based privacy-preserving system using events such as WINDOW CONTENT CHANGED, WINDOW STATE CHANGED, and VIEW SCROLLED was published for the first time in 2014~\cite{lau2014mimesis}. \revision{Further, NotificationListenerService API and US API (e.g., UsageStatsManager, UsageStats, UsageEvent, UsageEvents.Event) were released in API 18 and 21~\cite{NotificationListenerService, UsageStatsManager, UsageEvents, UsageEvents.Event}}. 

\emph{2015--2017: Notification APIs Updates \& Start Publishing Research on Various Themes}: \revision{From 2015 to 2017, there was no functional update to the AS API, and only continuous updates of NotificationListenerService and NotificationManager occurred as depicted in Figure \ref{fig:historyAPI}.} Nine studies (GAT: 3, CT: 1, APS: 2, and TRSGT: 3) corresponding to the PTS theme, a new theme was published using AS API~\cite{arruda2016capture, san2016gui, muangsiri2017random, wu2017appcheck, li2017programming, li2017sugilite, fazzini2017barista, liu2017mechanism}. Furthermore, NotificationListenerService API released in API level 18 in 2013 was first described as being utilized in a published paper corresponding to the purpose of NT/NM sub-theme of NP theme in 2017~\cite{pradhan2017understanding}; this work used the AS API and NotificationListenerService together. NotificationListenerService was used to track more abundant notification information (e.g., post, clear, action, title, id, style, modality), and information such as notification shading or duration was used AS API.

\emph{2018--2020: Research using US APIs on the Rise}: \revision{In 2018--2020, the window change-related events of AS API and the US API events (e.g., standby bucket changed/active/frequency, keyguard hidden/shown, screen interactive) updated in API 28. In 2018, 13 studies corresponding to EUI/UX theme were published.} \revision{This increment may be because the window changed related events of AS API was updated in Android 9 (released on March 7, 2018). There were also four papers on computational enhancement (CE), which aims to improve multi-modal interaction technologies such as Google Assistant (released in 2016).} \revision{In addition, studies using US APIs were on the rise as well.} \revision{However, these updated events of US API were not specifically used in the papers in 2019–2020.} As for those papers using the US API released in API 21 in 2014, two papers (MLMS: 1, GUPA: 1) were first published in 2016~\cite{welke2016differentiating, dutta2016introducing}, and US API was used in a total of five papers (NT: 1, EUI/UX: 1, UP: 2, PS: 1) in 2017–2018~\cite{ryu2017siginterface, singh2017usage, mehrotra2017understanding, qin2018deciphering, bonne2017exploring}. The number of studies using US API was eight papers (UP: 3, NT: 1, EUI/UX: 1, PS: 3) in 2019~\cite{samonte2019braille3d, qin2019association, yuan2019much, yan2019bridging, lee2019pass, torres2019behavioral, zhu2019riskcog, wang2019dcdroid} and six papers (UP: 2, PS: 3, PTS: 1) in 2020~\cite{radesky2020young, khan2020personal, zhu2020hybrid, andriotis2020allow, wang2020identifying, xiang2020dynamical}, a significant increase compared to 2015–2018. \revision{Because in 2018--2019, there were major updates of UsageEvent.Event (e.g., activity resumed/paused/stopped, device shut down, device start-up, and foreground service start/stop)~\cite{UsageEvents.Event} that can track app foreground, background status, and device/system status.} In 2020, there were only three studies (EUI/UX: 1, PS: 1, PTS: 1) using AS API~\cite{alshayban2020accessibility, leguesse2020reducing, li2020interactive}, which was far less compared to studies using US API. From 2016 to 2020, out of a total of 20 studies using US API, 9 were in the UP theme, accounting for 45 \% of the total. Therefore, the US API has been used in various sub-themes as a representative UP theme from 2015 to 2020. The current trend is increasingly using the US API by replacing the AS API for app usage log collection. 

\revision{\emph{Latest Trends as of January 2021–March 2022}: Note that our preliminary search of recent papers published in 2021/01--2022/03 (via ACM Digital Library, Scopus, Web of Science, and ScienceDirect) revealed that research themes of EUI/UX and Privacy \& Security are on the rise (12 EUI/UX, 9 PS, 4 PTS, 3 UP, and 2 NT studies) as depicted in Appendix D. In API level 32 of 2021, new events (e.g., SPEECH\_STATE\_LISTENING/SPEAKING/CHANGED\_START/END) associated with microphone’s Listening and Speaking status were added~\cite{AccessibilityEvent} and expected new studies through these events.}

\emph{Challenges of US/AS APIs}: App usage patterns can be grasped through the US API from 2015 onwards, making it more possible to understand various status information about app usage intuitively. Nevertheless, we cannot collect the information regarding interaction types (e.g., click, long click, scroll, focused, and typing) and interaction targets (UI elements and hierarchy) from the US API. In particular, studies corresponding to specific research purposes (PTS, PS) that require view information (interaction type, interaction target) inside the app had no choice but to rely on the AS API, which has an important function. While AS API can broaden the functionality of applications, it can potentially generate security risks. Once granted the user’s permissions, the API can be used to read data from other apps.  Therefore, Google currently regulates the AS API use. It can be used only to improve the accessibility for users with disabilities for apps registered in the Google Play Store\footnote{https://play.google.com/store}. Suppose the developer does not follow the policy. In that case, the policy support team of the Google Play Console\footnote{https://support.google.com/googleplay/android-developer/answer/7218994?hl=ko} will inform the developers to explain how the AS API is used in their developed application to help users with disabilities. If the third-party app does not meet the requirements within 30 days, the ability to AS API within the app will be removed, or the app is regulated and unpublished~\cite{androidpolice}. So far, for apps that use AS API other than those intended to assist the users with disabilities, there are no restrictions on publishing apps to other app stores or web hosting services (e.g., GitHub). \revision{Besides, when the AS API is used for the mobile usage and sensor data-driven research, there are no specific prohibition regulations by Google.} However, since the AS API has powerful functions that can fetch almost all user interaction data (e.g., interaction type/target and app status) about the app being used, the Google Android policy regulation may become more severe due to personal privacy and security issues. \revision{Hence,} even when using AS API for future research purposes, other alternatives are needed to prepare future restrictions.

\revision{\subsection{Privacy Issues in Personal Data Research}}
\label{sec:privacy}
In the surveyed studies of this article, mobile usage and sensor data collected and utilized through Android AS/US API and other APIs may cause privacy issues in the process of collection, storage, and utilization. For example, those studies that analyze smartphone data in the wild to understand users’ smartphone usage patterns and their surrounding contexts (i.e., interaction, context, and system sensing data) can be used to reconstruct a user’s everyday life patterns. Especially, AS API can be used to hijack sensitive personal data such as credit information~\cite{diao2019kindness}. Furthermore, among the surveyed studies of this article, studies that collect large-scale data over a long period may further increase the risk of privacy and security. Among the UP and NT research themes, if the goal is to identify the predictive features or build machine learning models, it would be beneficial to collect various types of smartphone data, possibly with many participants for a longer duration to improve the external validity of the research.

\revision{When conducting mobile usage and sensor data-driven research covered in this article, minimizing privacy risks is required,} and this should be carefully addressed in the user consent such as Institutional Review Board (IRB) documents and the app's request for consent. Indeed, surveyed studies in which data were collected from multiple participants over a long period of time in \revision{the wild} at risk for privacy were IRB-approved and users’ informed consent was obtained prior to the experiment~\cite{kim2019understanding, ferreira2015aware, welke2016differentiating, yuan2019much}. \revision{However, many studies do not mention whether the IRB-approved or users’ informed consent.} In addition, personal data was anonymized by removing personal details, only collecting universal unique ID (UUID), hashing sensitive information to minimize privacy and security risks~\cite{lee2018click, andone2016menthal}. Additionally, the collected data were stored in a local device rather than uploaded to the cloud service~\cite{bitsch2015psychologist}. Furthermore, researchers can follow well-known data protection guidelines principles that must be followed when using personal and sensitive user data. Appendix B offers the guidelines that researchers should follow when collecting and utilizing data to solve these privacy and data quality issues by referring to existing data protection guidelines (e.g., Google Play Console's Policy Center, U.S. Federal Trade Commission's fair information practices principles, EU General Data Protection Regulation).

\subsection{Data Quality Issues in Mobile Usage and Sensor Data-Driven Research}
\label{sec:quality}
Another challenge in studies of mobile usage and sensor data-driven analytics is data quality problems (e.g., missing value, outlier). Among the existing reviewed studies, we further investigated the quality issues that occurred during mobile usage and sensor data collection in the UP theme~\cite{lee2014hooked, singh2017usage, kim2019understanding, radesky2020young, khan2020personal, qin2019association, ferreira2015aware}. Major quality issues were categorized into human factors and system factors, and several guidelines for quality assurance are discussed.
In a long-term wild study, the duration of a participant’s data fails to meet the minimum period, and missing values occur possibly due to participants' personal situations or characteristics. Missing values occurred when participants did not follow the experimental guidelines; e.g., turning off wifi or GPS, turning off the device, and changing the smartphones during the study period \cite{singh2017usage, kim2019understanding, lee2014hooked}.

Data quality issues also arise due to system factors. Radesky et al. (2020) removed 13\% of the collected data not properly collected due to server and app usage logger’s problems (e.g., server maintenance, not installed necessary program for data collection)~\cite{radesky2020young}. In addition, Khan et al. (2020) mentioned several data quality issues 1) When tracking events that occur in smartphones through the app usage logger, data collection was impossible because Android built-in APIs were not supported in certain smartphone models. 2) Data accuracy (e.g., noise, outlier) deteriorated due to the poor quality of the sensor. 3) When the app usage logger app was installed, the battery drained quickly and the smartphone turned off easily, causing many missing values~\cite{khan2020personal}.  
Several papers mentioned strategies on data quality assurance. For example, researchers selected the participants based on their smartphone model and Android OS version. Khan et al. (2020) selected participants with the most suitable manufacturer (e.g., Samsung) in consideration of factors to improve data quality (e.g., sensor quality of smartphones, Android-APIs support for collected data, and battery persistence)~\cite{khan2020personal}. Human factor issues can be addressed by carefully setting up the experimental protocols and offering participant guidelines (e.g., use of smartphones for a certain period of time, system and configuration settings to always collect data). \revision{In addition, it is recommended to periodically check whether the data are collected~\cite{kim2019understanding}.}

\subsection{Limitation and Future Work}
\label{sec:limitation}
This article only surveyed the studies using APIs that collect smartphone data in Android \revision{environment}. Most studies based on mobile usage and sensor data-driven analytics used Android OS excepted research led by Apple Inc~\cite{chen2019developing}. According to Nishiyama et al.~\cite{nishiyama2020ios}, Android can flexibly distribute third-party apps that can access various sensors and collect mobile usage and sensor data. While iOS has limitations on accessing sensors and distributing applications. \revision{Because} iOS has limitations in sensing in the background, it supports that in only a few conditions (e.g., location updates, Bluetooth operations, background fetch, audio, and remote notifications). \revision{In addition, the difference between iOS and Android lies in the data types that can be collected.} \revision{On iOS, it is not easy to use APIs to track app usage patterns, touch interaction types, and UI components/hierarchies.} Access to mobile sensor data such as location, Bluetooth, telephony, and activity recognition in iOS must be justified for approval; iOS does not support accessing telephony, light, and temperature data~\cite{nishiyama2020ios}. Due to these limitations, prior research mostly used Android rather than iOS. \revision{Nevertheless, the global mobile operating system share of iOS devices in 2021 was about 17.2$\pm$4.2\% (calculated by the mean and standard deviation of 1st to 4th quarter in 2021)~\cite{Statista:iOSmarketshare}. Therefore, future survey research need to investigate mobile usage and sensor data-driven studies in the iOS environment.}

Furthermore, although we have separately identified the papers that entailed user-studies, this review did not investigate the experiment design such as the number of participants, experiment duration, laboratory/field testing, questionnaires data (including experience sampling method data) for IP1 studies. In future research, A review of experiment design and procedure (e.g., participant, duration, metric, tool) will help researchers when performing user studies for different purposes. Moreover, a follow-up review needs to analyze data analysis techniques when performing mobile usage and sensor data-driven analytics. For example, recent studies on digital phenotype extensively used mobile and sensor data to automatically classify user emotions and diseases.

\revision{Mobile context data-driven research in the Android OS} can be investigated using keywords different from the ones used in this article. \revision{To understand the personal daily activity logs, built-in sensors (e.g., Wi-Fi, GPS) and other APIs (e.g., Google Activity Recognition, \emph{SensorManager}) which are frequently used to collect user context data can be included as keywords.} \revision{The studies using} the AS/US APIs were investigated to find studies that effectively utilized app usage patterns. However, although the app usage pattern was collected in the Android \revision{OS}, studies that used their developed API names instead of names of describing related to the AS or US API were excluded from the review such as References~\cite{falaki2010diversity, banovic2014proactivetasks, ferreira2015securacy, shin2012understanding, abdullah2014towards, asselbergs2016mobile, chan2018students, eskes2016sociability, stutz2015smartphone, wang2014studentlife}. Therefore, it is expected that further, broader insights can be obtained by subsequent researchers if studies that collected the app usage patterns, despite not describing AS/US APIs in their papers, are included in the review scope.

\revision{The current work fails to capture the latest publications (January 2021--March 2022) because this research was conducted in the same year. An additional search at ACM Digital Library, Scopus, Web of Science, and ScienceDirect through the keywords in Table \ref{table:keywords} revealed 85 papers. Applying the PRISMA criteria in Figure \ref{fig:Flowchart} resulted in 30 papers, mostly published at well-known venues (e.g., ACM CHI/UIST/CCS, ASSET, PACM IMWUT/HCI, and USENIX S\&P). The category analysis results were discussed in Section \ref{sec:APIhistory} and the detailed statistics were reported in Appendix D.}

%Google Scholar was not considered an additional search DB, because Google Scholar’s results are overlapped with the other results of DB in major venues, and searched many studies non-peer-reviewed. 

%% file: 08conclusion.tex
\section{Conclusion}
\label{sec:conclusion}
This study analyzed prior studies that employed Android AS/US APIs to collect mobile usage and sensor data, which are the most representative forms of app usage data that can identify personal lifelog data. \revision{Through research purpose classification with five themes and 21 sub-themes,} this analysis helps researchers to understand at a glance the existing studies conducted for the past 10 years for each specific research purpose. \revision{In addition, this study presented standardized terms and taxonomy with a four-layer hierarchical structure from 109 studies reviewed, and this taxonomy helps to improve reproducibility for researchers and to better understand what data are used in the previous studies.} Furthermore, the tendency analysis of data types collected for each research purpose provides insight into reducing the cost and time for data collection and analysis. This study offers systematic guidelines for future research on Android AS/US APIs and lays a foundation for promoting research reproducibility and facilitating follow-up research on data-driven analytics with mobile \revision{usage and sensor} data.

%% file: 09appendix.tex
\section{appendix}
As the supplementary materials, we  discuss  the trends in research purpose and used data via IP types in Appendix A, and  the privacy and security issues in mobile usage and sensor data analytics based research in Appendix B. Appendix C describe AS/US API's events. \revision{Appendix D describes the API release and update via API level (Table A5) and categorized results of 30 papers published in January 2021–March 2022.} Further, a table of abbreviations was provided in Appendix E.